\documentclass[aps,prd,floats,floatfix,superscriptaddress,nofootinbib]{revtex4-1}
\usepackage{graphicx}
\usepackage{amsmath}
\usepackage{amssymb}
\usepackage{hyperref}
\usepackage{bbold}
\usepackage[normalem]{ulem}

\usepackage{color}

\begin{document}

\title{All the generalized Georgi-Machacek models}

\author{Heather E.~Logan}
\email{logan@physics.carleton.ca}

\affiliation{Ottawa-Carleton Institute for Physics, Carleton University, 1125 Colonel By Drive, Ottawa, Ontario K1S 5B6, Canada}

\author{Vikram Rentala}
\email{vikram.rentala@gmail.com}

\affiliation{Ottawa-Carleton Institute for Physics, Carleton University, 1125 Colonel By Drive, Ottawa, Ontario K1S 5B6, Canada}
\affiliation{School of Physical Sciences, Jawaharlal Nehru University, New Mehrauli Road, New Delhi - 110067, India}
\altaffiliation[Current address:]{Department of Physics, Indian Institute of Technology, Bombay, Mumbai - 400076, India}

\date{February 4, 2015}                                  

\begin{abstract}
The Georgi-Machacek model adds two SU(2)$_L$-triplet scalars to the Standard Model in such a way as to preserve custodial SU(2) symmetry.  We study the generalizations of the Georgi-Machacek model to SU(2)$_L$ representations larger than triplets.  Perturbative unitarity considerations limit the possibilities to models containing only SU(2)$_L$ quartets, quintets, or sextets.  These models are phenomenologically interesting because they allow the couplings of the 125~GeV Higgs boson to $WW$ and $ZZ$ to be larger than their values in the Standard Model.  We write down the most general custodial SU(2)-preserving scalar potentials for these models and outline their phenomenology.  We find that experimental and theoretical constraints on the fermiophobic custodial-fiveplet states present in each of the models lead to absolute upper bounds on the 125~GeV Higgs boson coupling strength to $WW$ and $ZZ$.
\end{abstract}

\maketitle

\section{Introduction}

Since the discovery of a Standard Model (SM)-like Higgs boson at the CERN Large Hadron Collider (LHC)~\cite{Aad:2012tfa}, there has been an increased interest in models with extended Higgs sectors to be used as benchmarks for LHC searches for physics beyond the SM.  One such model is the Georgi-Machacek (GM) model~\cite{Georgi:1985nv,Chanowitz:1985ug}, which adds isospin-triplet scalar fields to the SM in a way that preserves custodial SU(2) symmetry.  Its phenomenology has been extensively studied~\cite{Gunion:1989ci,Gunion:1990dt,HHG,Haber:1999zh,Aoki:2007ah,Godfrey:2010qb,Low:2010jp,Logan:2010en,Falkowski:2012vh,Chang:2012gn,Chiang:2012cn,Chiang:2013rua,Kanemura:2013mc,Englert:2013zpa,Killick:2013mya,Englert:2013wga,Efrati:2014uta,Hartling:2014zca,Chiang:2014hia,Chiang:2014bia,Godunov:2014waa,Hartling:2014aga,Chiang:2015kka}.  The GM model has also been incorporated into the scalar sectors of little Higgs~\cite{Chang:2003un,Chang:2003zn} and supersymmetric~\cite{Cort:2013foa,Garcia-Pepin:2014yfa} models, and an extension with an additional isospin doublet~\cite{Hedri:2013wea} has also been considered.

An interesting feature of the GM model that distinguishes it from extended Higgs sectors containing only isospin doublets and/or singlets is that the couplings of the SM-like Higgs boson to $WW$ and to $ZZ$ can be larger than in the SM.  Such an enhancement can also occur in an extension of the Higgs sector by an isospin septet with appropriately-chosen hypercharge~\cite{Hisano:2013sn,Kanemura:2013mc,Alvarado:2014jva,Geng:2014oea}.  These models are useful because they allow for a concrete study of the ``flat direction''~\cite{Zeppenfeld:2000td} that arises in the extraction of Higgs couplings from LHC data.  In particular, the on-resonance Higgs signal rate in a given production and decay channel can be written as
\begin{equation}
	{\rm Rate}_{ij} = \sigma_i \frac{\Gamma_j}{\Gamma_{\rm tot}}
	= \kappa_i^2 \sigma_i^{\rm SM} \frac{\kappa_j^2 \Gamma_j^{\rm SM}}
	{\sum_k \kappa_k^2 \Gamma_k^{\rm SM} + \Gamma_{\rm new}},
\end{equation}
where $\sigma_i$ is the Higgs production cross section in production mode $i$, $\Gamma_j$ is the Higgs decay partial width into final state $j$, $\Gamma_{\rm tot}$ is the total width of the Higgs boson, the corresponding quantities in the SM are denoted with a superscript, and $\Gamma_{\rm new}$ represents the partial width of the Higgs boson into new, non-SM final states.
One can then imagine a scenario in which all the coupling modification factors have a common value $\kappa_i \equiv \kappa > 1$ and there is a new, unobserved contribution to the Higgs total width, ${\rm BR}_{\rm new} > 0$.  In this case the Higgs production and decay rates measurable at the LHC are given by
\begin{equation}
	{\rm Rate}_{ij} = \frac{\kappa^4 \sigma_i^{\rm SM} \Gamma_j^{\rm SM}}{\kappa^2 \Gamma_{\rm tot}^{\rm SM} + \Gamma_{\rm new}}.
\end{equation}
All measured Higgs production and decay rates will be equal to their SM values if
\begin{equation}
	\kappa^2 = \frac{1}{1 - {\rm BR}_{\rm new}}, \qquad \qquad
	{\rm where} \quad
	{\rm BR}_{\rm new} \equiv \frac{\Gamma_{\rm new}}{\Gamma_{\rm tot}}
	= \frac{\Gamma_{\rm new}}{\kappa^2 \Gamma_{\rm tot}^{\rm SM} + \Gamma_{\rm new}}.
\end{equation}
In particular, a simultaneous enhancement of all the Higgs couplings to SM particles can mask, and be masked by, the presence of new decay modes of the Higgs that are not directly detected at the LHC.\footnote{Measuring such an enhancement in the Higgs couplings would be straightforward at a lepton-collider Higgs factory such as the International Linear Collider (ILC), where a direct measurement of the total Higgs production cross section in $e^+e^- \to Zh$ can be made with no reference to the Higgs decay branching ratios by using the recoil mass method (see, e.g., Ref.~\cite{Baer:2013cma}).}

One way to constrain these scenarios would be to constrain the total width of the Higgs boson at the LHC, for example through measurements of the off-shell production cross section of $gg \ (\to h^*) \to ZZ$~\cite{Kauer:2012hd,Caola:2013yja,Campbell:2013una}.  However, this measurement can become insensitive to a Higgs width enhancement if there are additional light scalars that contribute to the $gg \to ZZ$ process~\cite{Logan:2014ppa}, which is a generic feature of the models we study here.  This motivates the study of benchmark models in which an enhancement of the Higgs couplings to $WW$ and to $ZZ$ can be realized, in order to develop phenomenological strategies to constrain the enhanced-coupling scenario.

The GM model and the septet model mentioned above are the only two extended Higgs models currently on the market in which such an enhancement can be realized.  Both require an ultraviolet (UV) completion at scales not too much higher than the weak scale.  The custodial symmetry imposed on the scalar sector of the GM model is explicitly broken by hypercharge interactions~\cite{Gunion:1990dt,Englert:2013wga,Garcia-Pepin:2014yfa}, which implies that the custodial symmetry can only be exact at one energy scale.  This scale cannot be much higher than the weak scale~\cite{Garcia-Pepin:2014yfa}.  Similarly, in the septet model the septet must obtain its vacuum expectation value (vev) through a dimension-seven coupling to the SM Higgs doublet.  An explicit UV completion involving additional scalar fields was presented in Ref.~\cite{Hisano:2013sn}, but these new fields cannot be much heavier than the weak scale if a non-negligible septet vev is to be generated.  Despite these theoretical disadvantages, these models provide valuable phenomenological insight that cannot be obtained from Higgs sector extensions involving only isospin doublets and/or singlets.

It has long been known that the GM model can be generalized to include scalars in isospin representations larger than triplets, while maintaining custodial SU(2) invariance in the scalar potential~\cite{Galison:1983qg,Robinett:1985ec,Logan:1999if,Chang:2012gn}.  Though they suffer from the same hard breaking of custodial SU(2) symmetry by hypercharge gauge interactions as in the original GM model, such generalizations are phenomenologically interesting because they can accommodate even larger enhancements of the Higgs couplings to $WW$ and to $ZZ$ than in the original GM model.
In this paper we write down all such generalizations.  We start in Sec.~\ref{sec:framework} by reviewing the main features of the original GM model and how it can be generalized to higher isospin.  In Sec.~\ref{sec:unitarity} we determine the limit on the maximum isospin that is acceptable based on requiring perturbative unitarity in $2 \to 2$ scattering processes involving scalars and \emph{transverse} SU(2)$_L$ gauge bosons, following Ref.~\cite{Hally:2012pu}.  This limits us to only three generalizations of the GM model, which contain isospin quartets, quintets, or sextets.  In Sec.~\ref{sec:pheno} we outline the phenomenology of these three models and apply those experimental constraints that can be adapted from existing analyses in the GM model and others.  In Sec.~\ref{sec:potentials} we write down the most general scalar potentials for these three models, subject to the requirement that custodial symmetry is preserved, and give explicit formulas for the physical masses in terms of the parameters of the potentials.  We also comment on the decoupling behavior of the models.  We conclude in Sec.~\ref{sec:conclusions}.  In the Appendices we collect the SU(2) generators for higher isospin representations as well as the explicit expressions for the custodial-symmetry eigenstates in each of the models.

\section{Georgi-Machacek framework}
\label{sec:framework}

The SM Higgs sector possesses an accidental global SU(2)$_L \times$SU(2)$_R$ symmetry, where the SU(2)$_L$ is gauged to become the usual weak isospin gauge symmetry and the third generator of SU(2)$_R$ is gauged to become hypercharge (up to a normalization).  When electroweak symmetry is broken, the global SU(2)$_L \times$SU(2)$_R$ breaks down to its diagonal SU(2) subgroup, which is known as the custodial SU(2) symmetry.  The exact custodial symmetry in the SM has a slight explicit breaking due to the gauging of hypercharge and the difference of the top and bottom Yukawa couplings.  The Goldstone bosons transform as a custodial triplet, ensuring $M_{W^{\pm}} = M_{W^3}$ in the limit $g^{\prime} \to 0$.  This leads to the well-known result $\rho \equiv M_W^2 / M_Z^2 \cos^2\theta_W = 1$ at tree level.

The scalar sector of the GM model~\cite{Georgi:1985nv,Chanowitz:1985ug} consists of the usual complex doublet $(\phi^+,\phi^0)$ with hypercharge\footnote{We use the convention $Q = T^3 + Y/2$ to define the hypercharge normalization.} $Y = 1$, a real triplet $(\xi^+,\xi^0,\xi^-)$ with $Y = 0$, and  a complex triplet $(\chi^{++},\chi^+,\chi^0)$ with $Y=2$.  With this field content, the entire scalar sector can be made invariant under the global SU(2)$_L \times$SU(2)$_R$ symmetry, thereby preserving custodial SU(2) in the scalar sector after electroweak symmetry breaking.  The doublet is responsible for the fermion masses as in the SM.

In order to make the global SU(2)$_L \times$SU(2)$_R$ symmetry explicit, we write the doublet in the form of a bidoublet $\Phi$ and combine the triplets to form a bitriplet $X$:
\begin{equation}
	\Phi = \left( \begin{array}{cc}
	\phi^{0*} &\phi^+  \\
	-\phi^{+*} & \phi^0  \end{array} \right), \qquad
	X =
	\left(
	\begin{array}{ccc}
	\chi^{0*} & \xi^+ & \chi^{++} \\
	 -\chi^{+*} & \xi^{0} & \chi^+ \\
	 \chi^{++*} & -\xi^{+*} & \chi^0
	\end{array}
	\right).
	\label{eq:PX}
\end{equation}
The vevs in the electroweak symmetry breaking vacuum are defined by $\langle \Phi  \rangle = \frac{ v_{\phi}}{\sqrt{2}} \mathbb{1}_{2\times2}$  and $\langle X \rangle = v_{\chi} \mathbb{1}_{3 \times 3}$, where the $W$ and $Z$ boson masses constrain
\begin{equation}
	v_{\phi}^2 + 8 v_{\chi}^2 \equiv v^2 = \frac{1}{\sqrt{2} G_F} \approx (246~{\rm GeV})^2.
	\label{eq:vevrelation}
\end{equation}

The most general gauge-invariant scalar potential involving these fields that conserves custodial SU(2) is given, in the conventions of Ref.~\cite{Hartling:2014zca}, by\footnote{A translation table to other parameterizations in the literature has been given in the appendix of Ref.~\cite{Hartling:2014zca}.}
\begin{eqnarray}
	V(\Phi,X) &= & \frac{\mu_2^2}{2}  \text{Tr}(\Phi^\dagger \Phi)
	+  \frac{\mu_3^2}{2}  \text{Tr}(X^\dagger X)
	+ \lambda_1 [\text{Tr}(\Phi^\dagger \Phi)]^2
	+ \lambda_2 \text{Tr}(\Phi^\dagger \Phi) \text{Tr}(X^\dagger X)   \nonumber \\
          & & + \lambda_3 \text{Tr}(X^\dagger X X^\dagger X)
          + \lambda_4 [\text{Tr}(X^\dagger X)]^2
           - \lambda_5 \text{Tr}( \Phi^\dagger \tau^a \Phi \tau^b) \text{Tr}( X^\dagger T^a_1 X T^b_1)
           \nonumber \\
           & & - M_1 \text{Tr}(\Phi^\dagger \tau^a \Phi \tau^b)(U X U^\dagger)_{ab}
           -  M_2 \text{Tr}(X^\dagger T^a_1 X T^b_1)(U X U^\dagger)_{ab}.
           \label{eq:potential}
\end{eqnarray}
Here the SU(2) generators for the doublet representation are $\tau^a = \sigma^a/2$ with $\sigma^a$ being the Pauli matrices and the generators for the triplet representation $T^a_1$ are given in Appendix~\ref{app:generators}.
The matrix $U$, which rotates $X$ into the Cartesian basis, is given by~\cite{Aoki:2007ah}
\begin{equation}
	 U = \left( \begin{array}{ccc}
	- \frac{1}{\sqrt{2}} & 0 &  \frac{1}{\sqrt{2}} \\
	 - \frac{i}{\sqrt{2}} & 0  &   - \frac{i}{\sqrt{2}} \\
	   0 & 1 & 0 \end{array} \right).
	 \label{eq:U}
\end{equation}
Alternatively, the two trilinear terms can be rewritten as
\begin{eqnarray}
	{\rm Tr}(\Phi^\dagger \tau^a \Phi \tau^b)(U X U^\dagger)_{ab}
	&=& {\rm Tr} \left[\Phi^{\dagger} \widehat{T}^{1,i}_{1/2} \Phi (\widehat{T}^{1,j}_{1/2})^\dagger \right] X_{ij}, \nonumber \\
	{\rm Tr}(X^\dagger T^a_1 X T^b_1)(U X U^\dagger)_{ab}
           &=& {\rm Tr }\left[ X^\dagger \widehat{T}^{1,i}_{1} X (\widehat{T}^{1,j}_{1})^\dagger \right] X_{ij},
\end{eqnarray}
where we use the notation $\widehat{T}^{j,i}_r$ to denote the $i$-th spherical tensor of rank $j$ constructed from the basis of generators $(-\frac{1}{\sqrt{2}} T^+_r , T^3_r, \frac{1}{\sqrt{2}} T^-_r)$ in representation $r$.  Higher rank tensors are constructed via tensor products of the rank-1 SU(2) generators. Here $i$ runs from $j$ to $-j$ in integer steps and corresponds to the indices of $X$, which is naturally defined in the spherical basis as in Eq.~(\ref{eq:PX}).  Explicit expressions for the spherical tensors are given in Appendix~\ref{app:spherical}.

The physical fields can be organized by their transformation properties under the custodial SU(2) symmetry into a fiveplet, a triplet, and two singlets.  The fiveplet and triplet states are given by\footnote{For consistency with our construction of the custodial fiveplet of the generalized GM models, we have adopted the opposite sign convention for $H_5^0$ compared to that in, e.g., Refs.~\cite{Gunion:1989ci,HHG,Hartling:2014zca}.  This leads to an overall minus sign in the $H_5^0 VV$ couplings in Eq.~(\ref{eq:H5VV}) compared to those in Refs.~\cite{Gunion:1989ci,HHG,Hartling:2014zca}, but has no physical consequences.  We apologize for contributing to the proliferation of conventions.}
\begin{eqnarray}
	&&H_5^{++} = \chi^{++}, \qquad
	H_5^+ = \frac{\left(\chi^+ - \xi^+\right)}{\sqrt{2}}, \qquad
	H_5^0 = -\sqrt{\frac{2}{3}} \xi^0 + \sqrt{\frac{1}{3}} \chi^{0,r}, \nonumber \\
	&&H_3^+ = - s_H \phi^+ + c_H \frac{\left(\chi^++\xi^+\right)}{\sqrt{2}}, \qquad
	H_3^0 = - s_H \phi^{0,i} + c_H \chi^{0,i},
\end{eqnarray}
where the vevs are parameterized by
\begin{equation}
	c_H \equiv \cos\theta_H = \frac{v_{\phi}}{v}, \qquad
	s_H \equiv \sin\theta_H = \frac{2\sqrt{2}\,v_\chi}{v},
\end{equation}
and we have decomposed the neutral fields into real and imaginary parts according to
\begin{equation}
	\phi^0 \to \frac{v_{\phi}}{\sqrt{2}} + \frac{\phi^{0,r} + i \phi^{0,i}}{\sqrt{2}},
	\qquad
	\chi^0 \to v_{\chi} + \frac{\chi^{0,r} + i \chi^{0,i}}{\sqrt{2}},
	\qquad
	\xi^0 \to v_{\chi} + \xi^0.
	\label{eq:decomposition}
\end{equation}
The masses within each custodial multiplet are degenerate at tree level and can be written (after eliminating $\mu_2^2$ and $\mu_3^2$ in favor of the vevs) as\footnote{Note that the ratio $M_1/v_{\chi}$ is finite in the limit $v_{\chi} \to 0$,
\begin{equation}
	\frac{M_1}{v_{\chi}} = \frac{4}{v_{\phi}^2}
	\left[ \mu_3^2 + (2 \lambda_2 - \lambda_5) v_{\phi}^2
	+ 4(\lambda_3 + 3 \lambda_4) v_{\chi}^2 - 6 M_2 v_{\chi} \right],
\end{equation}
which follows from the minimization condition $\partial V/\partial v_{\chi} = 0$.}
\begin{eqnarray}
	m_5^2 &=& \frac{M_1}{4 v_{\chi}} v_\phi^2 + 12 M_2 v_{\chi}
	+ \frac{3}{2} \lambda_5 v_{\phi}^2 + 8 \lambda_3 v_{\chi}^2, \nonumber \\
	m_3^2 &=&  \frac{M_1}{4 v_{\chi}} (v_\phi^2 + 8 v_{\chi}^2)
	+ \frac{\lambda_5}{2} (v_{\phi}^2 + 8 v_{\chi}^2)
	= \left(  \frac{M_1}{4 v_{\chi}} + \frac{\lambda_5}{2} \right) v^2.
\end{eqnarray}

The two custodial SU(2)--singlet mass eigenstates are given by
\begin{equation}
	h = \cos \alpha \, \phi^{0,r} - \sin \alpha \, H_1^{0\prime},  \qquad
	H = \sin \alpha \, \phi^{0,r} + \cos \alpha \, H_1^{0\prime},
	\label{mh-mH}
\end{equation}
where
\begin{equation}
	H_1^{0 \prime} = \sqrt{\frac{1}{3}} \xi^0 + \sqrt{\frac{2}{3}} \chi^{0,r}.
\end{equation}
The mixing angle and masses are given by
\begin{eqnarray}
	&&\sin 2 \alpha =  \frac{2 \mathcal{M}^2_{12}}{m_H^2 - m_h^2},    \qquad
	\cos 2 \alpha =  \frac{ \mathcal{M}^2_{22} - \mathcal{M}^2_{11}  }{m_H^2 - m_h^2},
	\nonumber \\
	&&m^2_{h,H} = \frac{1}{2} \left[ \mathcal{M}_{11}^2 + \mathcal{M}_{22}^2
	\mp \sqrt{\left( \mathcal{M}_{11}^2 - \mathcal{M}_{22}^2 \right)^2
	+ 4 \left( \mathcal{M}_{12}^2 \right)^2} \right],
	\label{eq:hmass}
\end{eqnarray}
where we choose $m_h < m_H$, and
\begin{eqnarray}
	\mathcal{M}_{11}^2 &=& 8 \lambda_1 v_{\phi}^2, \nonumber \\
	\mathcal{M}_{12}^2 &=& \frac{\sqrt{3}}{2} v_{\phi}
	\left[ - M_1 + 4 \left(2 \lambda_2 - \lambda_5 \right) v_{\chi} \right], \nonumber \\
	\mathcal{M}_{22}^2 &=& \frac{M_1 v_{\phi}^2}{4 v_{\chi}} - 6 M_2 v_{\chi}
	+ 8 \left( \lambda_3 + 3 \lambda_4 \right) v_{\chi}^2.
\end{eqnarray}

The GM model can be generalized in a straightforward way by replacing the bitriplet with a larger representation under SU(2)$_L\times$SU(2)$_R$~\cite{Galison:1983qg,Robinett:1985ec,Logan:1999if,Chang:2012gn}.  Because custodial symmetry is still preserved in the scalar sector, the physical states can still be classified according to their transformation properties under custodial SU(2); this leads to a variety of generic results~\cite{Logan:1999if} that can be expressed in terms of the isospin $(T,T)$ of the larger representation.  We will refer to these generalized Georgi-Machacek models by using the notation GGM$(2T+1)$.

\section{Constraints from perturbative unitarity}
\label{sec:unitarity}

Perturbative unitarity of tree-level $2 \to 2$ scattering amplitudes involving pairs of scalars and pairs of \emph{transversely} polarized SU(2)$_L$ gauge bosons limits the maximum isospin of the scalars.  The largest eigenvalue of the coupled-channel scattering matrix for such scattering involving a single complex scalar multiplet with isospin $T$ is given by~\cite{Hally:2012pu}
\begin{equation}
	a_{0,c}^{\rm max, SU(2)}(T) = \frac{g^2}{16 \pi} \frac{(n^2 - 1) \sqrt{n}}{2 \sqrt{3}},
\end{equation}
where $n \equiv 2T + 1$ is the number of states in the multiplet.  For a real multiplet, the eigenvalue is $a_{0,r}^{\rm max, SU(2)}(T) = a_{0,c}^{\rm max, SU(2)}(T)/\sqrt{2}$.

In a model with more than one scalar multiplet, the largest eigenvalue of the overall scattering matrix is found by adding the eigenvalues for each individual multiplet in quadrature.  (We ignore the contributions from scattering processes involving transversely polarized hypercharge gauge bosons; including them would not change our overall conclusions below.)

Results for the models of interest are summarized in Table~\ref{tab:uni}.  For the numerical calculation, we take $\alpha_{em} = s_W^2 g^2/4 \pi \simeq 1/128$ and $s_W^2 \simeq 0.231$.  We impose the perturbative unitarity constraint $|{\rm Re} \, a_0| < 1/2$.  This eliminates all generalized GM models containing septets or larger representations.

\begin{table}
\begin{tabular}{c c cc c c c}
\hline\hline
Model name & SU(2)$_L \times$SU(2)$_R$ reps & $T$ & $Y$ & real/complex & $a_0^{\rm max,SU(2)}$ & $b_2$ \\
\hline
GM & $(2 \times 2) + (3 \times 3)$ & 1/2 & 1 & complex & 0.043 & $-13/6$ \\
 & & 1 & 2 & complex & & \\
 & & 1 & 0 & real & & \\
\hline
GGM4 & $(2 \times 2) + (4 \times 4)$ & 1/2 & 1 & complex & 0.104 & 1/6 \\
 & & 3/2 & 3 & complex & & \\
 & & 3/2 & 1 & complex & & \\
\hline
GGM5 & $(2 \times 2) + (5 \times 5)$ & 1/2 & 1 & complex & 0.207 & 31/6 \\
 & & 2 & 4 & complex & & \\
 & & 2 & 2 & complex & & \\
 & & 2 & 0 & real & & \\
\hline
GGM6 & $(2 \times 2) + (6 \times 6)$ & 1/2 & 1 & complex & 0.363 & $43/3$ \\
 & & 5/2 & 5 & complex & & \\
 & & 5/2 & 3 & complex & & \\
 & & 5/2 & 1 & complex & & \\
\hline
GGM7 & $(2 \times 2) + (7 \times 7)$ & 1/2 & 1 & complex & 0.580 & $59/2$  \\
(excluded)  & & 3 & 6 & complex & & \\
 & & 3 & 4 & complex & & \\
 & & 3 & 2 & complex & & \\
 & & 3 & 0 & real & & \\
\hline\hline
\end{tabular}
\caption{Scalar field content and largest eigenvalue of the coupled-channel scattering matrix for scattering of pairs of transverse SU(2)$_L$ gauge bosons into pairs of scalars.  The GGM7 and higher models are excluded by the perturbative unitarity requirement $|{\rm Re} \, a_0| < 1/2$.  We also give the one-loop SU(2)$_L$ beta function coefficient including the contribution of the new scalars.  The $a_0$ values include the contributions from all scalar fields added in quadrature, including the doublet.}
\label{tab:uni}
\end{table}

We are left with only three models beyond the familiar GM model with triplets:
\begin{itemize}
\item[$(i)$] GGM4, containing two complex isospin-quartets in addition to the SM Higgs doublet;
\item[$(ii)$] GGM5, containing two complex isospin-quintets and one real isospin-quintet in addition to the SM Higgs doublet; and
\item[$(iii)$] GGM6, containing three complex isospin-sextets in addition to the SM Higgs doublet.
\end{itemize}

For completeness we also compute the one-loop SU(2)$_L$ beta function coefficient $b_2$ including the contributions of the additional scalars.  This is given by
\begin{equation}
	b_2 = -\frac{19}{6} + N \frac{n (n^2 - 1)}{36},
\end{equation}
where $-19/6$ is the SM contribution including the SM Higgs doublet, $n = 2T+1$ is the size of the additional multiplets, and $N$ is equal to the number of complex scalars of isospin $T$ plus half the number of real scalars of isospin $T$.  The value of $\alpha_2$ at scale $\mu$ is given in terms of the value at $M_Z$, $\alpha_2(M_Z) \equiv g^2/4 \pi$, by
\begin{equation}
	\alpha_2^{-1}(\mu) = \alpha_2^{-1}(M_Z) - \frac{b_2}{2 \pi} \log \left( \frac{\mu}{M_Z} \right).
\end{equation}
The value of $b_2$ for each of the models is given in Table~\ref{tab:uni}.

\section{Phenomenology}
\label{sec:pheno}

In this section we outline some of the phenomenological features of these models.  The results in this section can in fact be derived using the custodial symmetry, without reference to the explicit forms of the scalar potentials that will be given in the next section.

\subsection{Vevs and physical states}
\label{sec:states}

We start by defining the vevs of the bidoublet $\Phi$ and the $(n \times n)$ representation $X_n$ with isospin $T = (n-1)/2$ as
\begin{equation}
	\langle \Phi \rangle = \frac{v_{\phi}}{\sqrt{2}} \mathbb{1}_{2 \times 2}, \qquad \qquad
	\langle X_n \rangle = v_n \mathbb{1}_{n \times n}.
	\label{eq:vevs}
\end{equation}
We can choose the vevs to be positive without loss of generality.
The $W$ mass constrains these vevs according to~\cite{Robinett:1985ec,Logan:1999if}
\begin{equation}
	v_{\phi}^2 + \frac{4}{3} T (T+1) (2T+1) v_n^2 = v^2
	\equiv \frac{1}{\sqrt{2} G_F} \simeq (246~{\rm GeV})^2.
\end{equation}
For the GM model and its extensions, this corresponds to
\begin{eqnarray}
	{\rm GM}: &\qquad& v^2 = v_{\phi}^2 + 8 v_{\chi}^2, \nonumber \\
	{\rm GGM4}: &\qquad& v^2 = v_{\phi}^2 + 20 \, v_4^2, \nonumber \\
	{\rm GGM5}: &\qquad& v^2 = v_{\phi}^2 + 40 \, v_5^2, \nonumber \\
	{\rm GGM6}: &\qquad& v^2 = v_{\phi}^2 + 70 \, v_6^2.
\end{eqnarray}
In each case we define
\begin{equation}
	c_H \equiv \cos \theta_H = \frac{v_{\phi}}{v}.
\end{equation}
Then
\begin{equation}
	s_H \equiv \sin \theta_H = \left\{ \begin{array}{ll}
		\sqrt{8} \, v_{\chi}/v \qquad & {\rm GM} \\
		\sqrt{20} \, v_4/v & {\rm GGM4} \\
		\sqrt{40} \, v_5/v & {\rm GGM5} \\
		\sqrt{70} \, v_6/v & {\rm GGM6}.
		\end{array} \right.
\end{equation}

After electroweak symmetry breaking, the bidoublet and the $(n\times n)$ representation break down into multiplets of custodial SU(2) as follows:
\begin{eqnarray}
	\Phi: &\qquad & 2 \otimes 2 \to 3 \oplus 1 \nonumber \\
	X_3: &\qquad & 3 \otimes 3 \to 5 \oplus 3 \oplus 1 \nonumber \\
	X_4: &\qquad & 4 \otimes 4 \to 7 \oplus 5 \oplus 3 \oplus 1 \nonumber \\
	X_5: &\qquad & 5 \otimes 5 \to 9 \oplus 7 \oplus 5 \oplus 3 \oplus 1 \nonumber \\
	X_6: &\qquad & 6 \otimes 6 \to 11 \oplus 9 \oplus 7 \oplus 5 \oplus 3 \oplus 1.
\end{eqnarray}
Explicit expressions for all the custodial-symmetry eigenstates are given in Appendix~\ref{app:states}.

Defining $\Phi = (\phi^+, (v_{\phi} + \phi^{0,r} + i \phi^{0,i})/\sqrt{2})^T$, the custodial singlet in $\Phi$ is the state $\phi^{0,r}$ while the custodial triplet is $\Phi_3 \equiv (\Phi_3^+, i\Phi_3^0, \Phi_3^-)^T = (\phi^+, i\phi^{0,i}, \phi^{+*})^T$.  For each of these models, we will denote the custodial singlet in $X_n$ as $H_1^{\prime 0}$ and the custodial triplet as $H_3^{\prime} \equiv (H_3^{\prime +}, i H_3^{\prime 0}, H_3^{\prime -})^T$.  The primes indicate that these are not mass eigenstates.  The custodial fiveplet and higher representations do not mix and are mass eigenstates; we will denote these custodial multiplets as $H_5$, $H_7$, etc., with masses $m_5$, $m_7$, etc., respectively.

In each model, the custodial triplet from $\Phi$ mixes with the custodial triplet from $X_n$ to yield a triplet of Goldstone bosons which are eaten by the $W^\pm$ and $Z$ bosons, and a physical custodial triplet $H_3$.  In all the models these states are given by the expressions~\cite{Logan:1999if}
\begin{eqnarray}
	G^{0, \pm} &=& c_H \Phi_3^{0, \pm} + s_H H_3^{\prime 0, \pm}, \nonumber \\
	H_3^{0, \pm} &=& -s_H \Phi_3^{0, \pm} + c_H H_3^{\prime 0, \pm}.
\end{eqnarray}
We denote the mass of the physical custodial triplet by $m_3$.

The custodial singlets mix by an angle $\alpha$ to form mass eigenstates $h$ and $H$, defined so that $m_h < m_H$:
\begin{eqnarray}
	h &=& c_{\alpha} \phi^{0,r} - s_{\alpha} H_1^{\prime 0}, \nonumber \\
	H &=& s_{\alpha} \phi^{0,r} + c_{\alpha} H_1^{\prime 0},
\end{eqnarray}
where we use the shorthand notation $s_{\alpha} = \sin\alpha$ and $c_{\alpha} = \cos\alpha$.
The angle $\alpha$ is determined by the parameters of the scalar potential.

\subsection{Couplings}

Given these mixing angles, the couplings of all the scalar states to fermions can be defined.  Fermion masses are generated by the SU(2)$_L$ doublet in the same way as in the SM.  Because $h$, $H$, and $H_3$ are the only states that contain a doublet admixture, they are the only scalars that will couple to fermions; the rest of the states, $H_5$, $H_7$, etc., are fermiophobic.  The Feynman rules are identical to those in the GM model~\cite{Gunion:1989ci,HHG,Logan:1999if,Hartling:2014zca} (we use the sign convention of Ref.~\cite{Hartling:2014zca} for $H_3^0$):
\begin{eqnarray}
	h \bar f f: &\quad& -i \frac{m_f}{v} \frac{\cos \alpha}{\cos \theta_H} \equiv -i \frac{m_f}{v} \kappa_f^h, \qquad \qquad
	H \bar f f: \quad -i \frac{m_f}{v} \frac{\sin \alpha}{\cos \theta_H} \equiv -i \frac{m_f}{v} \kappa_f^H, \nonumber \\
	H_3^0 \bar u u: &\quad& \frac{m_u}{v} \tan \theta_H \gamma_5, \qquad \qquad \qquad \quad\qquad
	H_3^0 \bar d d: \quad -\frac{m_d}{v} \tan \theta_H \gamma_5, \nonumber \\
	H_3^+ \bar u d: &\quad& -i \frac{\sqrt{2}}{v} V_{ud} \tan\theta_H
		\left( m_u P_L - m_d P_R \right), \nonumber \\
	H_3^+ \bar \nu \ell: &\quad& i \frac{\sqrt{2}}{v} \tan\theta_H m_{\ell} P_R.
\end{eqnarray}
Here $f$ is any charged fermion, $V_{ud}$ is the appropriate element of the Cabibbo-Kobayashi-Maskawa (CKM) matrix, and the projection operators are defined as $P_{R,L} = (1 \pm \gamma_5)/2$.  The $H_3^0 \bar \ell \ell$ couplings are the same as the $H_3^0 \bar d d$ couplings with $m_d \to m_{\ell}$.

Custodial symmetry also fixes the coupling~\cite{Logan:1999if}
\begin{equation}
	Z_{\mu} H_3^+ H_3^-: \quad i \frac{e}{2 s_W c_W} (1 - 2 s_W^2) (p_+ - p_-)_{\mu},
\end{equation}
where $p_{\pm}$ are the incoming momenta of $H_3^{\pm}$, respectively, $s_W$ and $c_W$ denote the sine and cosine of the weak mixing angle, and the covariant derivative is given by
\begin{equation}
	\mathcal{D}_{\mu} = \partial_{\mu} - i \frac{g}{\sqrt{2}} \left( W_{\mu}^+ T^+ + W_{\mu}^- T^- \right) - i \frac{e}{s_W c_W} Z_{\mu} \left( T^3 - s_W^2 Q \right) - i e A_{\mu} Q.
\end{equation}

We note that, for all the generalized GM models, the couplings of $H_3^{\pm}$ to fermions and to the $Z$ boson are identical to the corresponding couplings of $H^{\pm}$ in the Type-I two Higgs doublet model~\cite{Branco:2011iw}, with the replacement $\cot\beta \to \tan\theta_H$.  This implies that the constraints on the $(m_3, s_H)$ plane in the GM model from $b \to s \gamma$~\cite{Hartling:2014aga} can be directly applied to all the generalized GM models.  We will illustrate this in the next subsection.

We now write down the couplings of $h$ and $H$ to vector boson pairs.  These can be written for all the generalized GM models as
\begin{eqnarray}
	\kappa_V^h &=& c_{\alpha} c_H - \sqrt{A} \, s_{\alpha} s_H, \nonumber \\
	\kappa_V^H &=& s_{\alpha} c_H + \sqrt{A} \, c_{\alpha} s_H,
	\label{eq:kappaV}
\end{eqnarray}
where~\cite{Logan:1999if}
\begin{equation}
	A = \frac{4}{3} T (T+1),
	\label{eq:A}
\end{equation}
and $\kappa_V^h$ is defined as the coupling of $h$ to $VV$ ($V = W$ or $Z$) normalized to its SM value, and similarly for $H$.  In what follows we will assume that $h$ is the discovered 125~GeV Higgs boson.  We see that the special case of simultaneous enhancement of the $h$ couplings to fermions and to vector bosons, $\kappa_f^h = \kappa_V^h$, is obtained when
\begin{equation}
	\frac{c_{\alpha}}{s_{\alpha}} = - \sqrt{A} \frac{c_H}{s_H}.
\end{equation}
To simultaneously obtain the same enhancement of the $h\gamma\gamma$ coupling requires that the sum of the contributions of the charged scalars to the loop-induced $h \to \gamma\gamma$ vertex vanishes.
In what follows we will not impose these requirements; instead we will examine the maximum possible enhancement of $\kappa_V^h$ allowed by constraints on the additional Higgs particles in the models and leave a full study of the constraints from the 125~GeV Higgs signal strength measurements to future work.

\begin{figure}
\resizebox{0.6\textwidth}{!}{\includegraphics{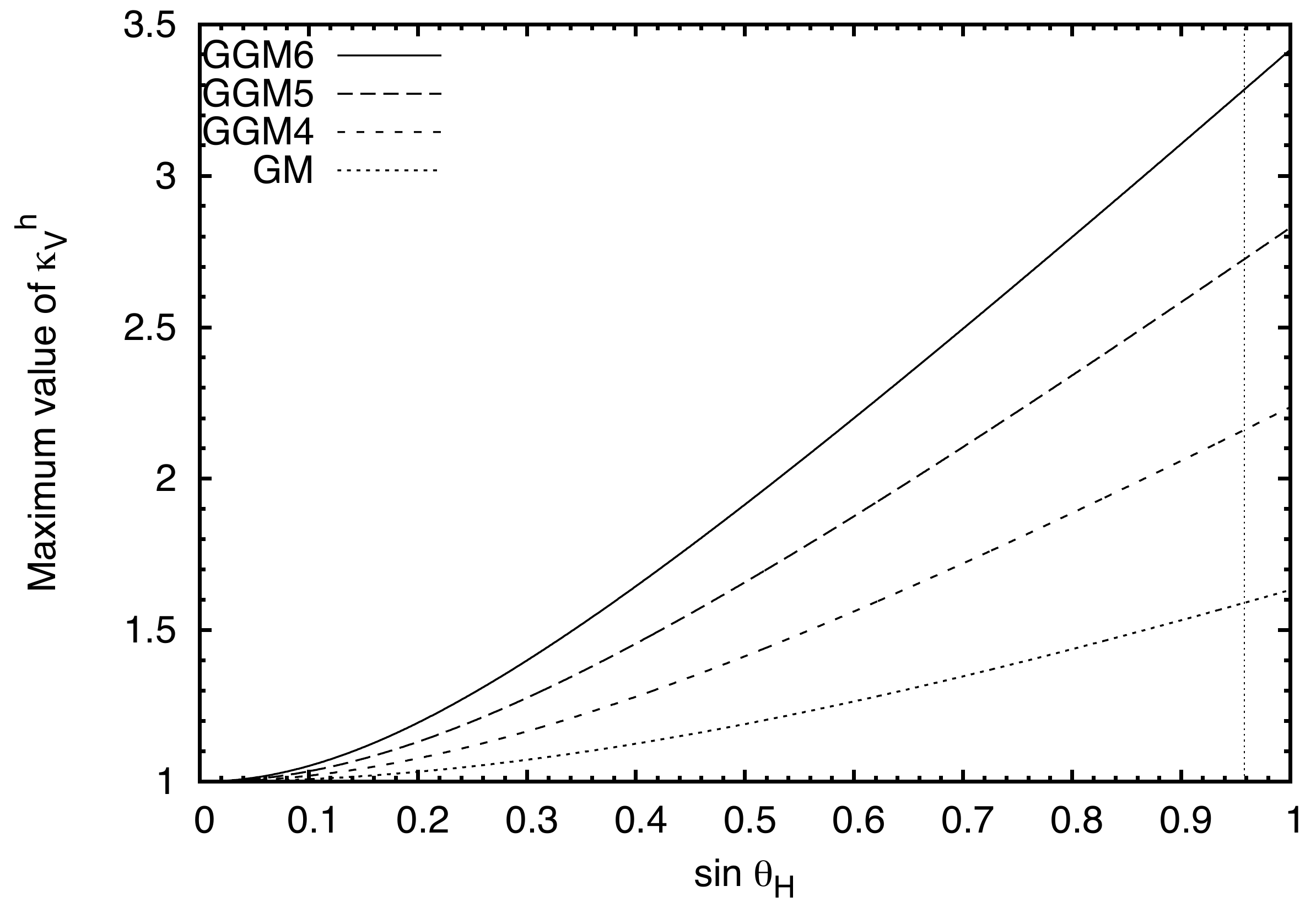}}
\caption{Maximum value of $\kappa_V^h$ as a function of $s_H$ in the GM model and the three generalized GM models.  The dotted vertical line indicates the limit $\tan\theta_H < 10/3$ imposed to avoid nonperturbative values of the top quark Yukawa coupling~\cite{Barger:1989fj}.}
\label{fig:sh-kmax}
\end{figure}

For the models under consideration we have
\begin{eqnarray}
	{\rm GM}: &\qquad& A = 8/3,  \nonumber \\
	{\rm GGM4}: &\qquad& A = 5, \nonumber \\
	{\rm GGM5}: &\qquad& A = 8, \nonumber \\
	{\rm GGM6}: &\qquad& A = 35/3.
	\label{eq:Avalues}
\end{eqnarray}
These lead to absolute upper bounds on $\kappa_V^h$ of
\begin{eqnarray}
	{\rm GM}: &\qquad& \kappa_V^h \leq \sqrt{8/3} \simeq 1.63,  \nonumber \\
	{\rm GGM4}: &\qquad& \kappa_V^h \leq \sqrt{5} \simeq 2.24, \nonumber \\
	{\rm GGM5}: &\qquad& \kappa_V^h \leq \sqrt{8} \simeq 2.83, \nonumber \\
	{\rm GGM6}: &\qquad& \kappa_V^h \leq \sqrt{35/3} \simeq 3.42.
	\label{eq:kappabounds}
\end{eqnarray}
These bounds are saturated when $s_H \to 1$, $s_{\alpha} \to -1$.  Such a limit cannot be obtained in practice because $s_H \to 1$ corresponds to $v_{\phi} \to 0$, in which case the fermion Yukawa couplings blow up.  To avoid parameter regions in which the top quark Yukawa coupling becomes too large, one should impose a lower bound on $v_{\phi}$; following the numerical choice made in Ref.~\cite{Barger:1989fj} yields $\tan\theta_H < 10/3$.  The upper bounds given in Eq.~(\ref{eq:kappabounds}) then become $1.59$, $2.16$, $2.72$, and $3.28$, respectively.
The upper bound on $\kappa_V^h$ as a function of $s_H$ in each model is shown in Fig.~\ref{fig:sh-kmax}, where we have chosen the value of $\alpha$ at each point that maximizes $\kappa_V^h$.\footnote{The value of $\alpha$ that maximizes $\kappa_V^h$ also yields $\kappa_V^H = 0$, so that this upper bound can also be found using
\begin{equation}
	\kappa_V^h \leq \left[ (\kappa_V^h)^2 + (\kappa_V^H)^2 \right]^{1/2}
		= \left[ 1 + (A-1) s_H^2 \right]^{1/2}.
\end{equation}}

In Ref.~\cite{Hartling:2014aga} it was shown that, once indirect experimental constraints (primarily from $b \to s\gamma$) and theoretical constraints on the scalar potential are taken into account, the upper bound on $\kappa_V^h$ in the GM model is further reduced to $1.36$.  We illustrate in the next subsection the effect of applying the indirect experimental constraint from $b \to s \gamma$.  However, a full treatment of the theoretical constraints on the generalized GM models is beyond the scope of this paper.

The couplings of the custodial fiveplet $H_5$ can be deduced in all the generalized GM models based on the requirement that the bad high-energy behavior of the longitudinal $VV \to VV$ scattering amplitudes is properly cancelled by scalar exchange, thereby restoring unitarity in the high-energy limit~\cite{HHG,Falkowski:2012vh,Grinstein:2013fia,Bellazzini:2014waa}.  In each of these models, the unitarization of the $VV \to VV$ amplitudes is accomplished through the exchange of $h$, $H$, and $H_5$, due to the preservation of custodial symmetry~\cite{Falkowski:2012vh,Grinstein:2013fia}.  Custodial symmetry forces the $H_5VV$ Feynman rules to take the form~\cite{Falkowski:2012vh}\footnote{Our sign conventions for $H_5^+$ and $H_5^0$ yield an extra minus sign in their Feynman rules compared to the corresponding expressions in Ref.~\cite{Falkowski:2012vh}.}
\begin{eqnarray}
	H_5^0 W^+_{\mu} W^-_{\nu}: &\quad & - i \frac{2 M_W^2}{v} \frac{g_5}{\sqrt{6}} g_{\mu\nu},
		\nonumber \\
	H_5^0 Z_{\mu} Z_{\nu}: &\quad & i \frac{2 M_Z^2}{v} \sqrt{\frac{2}{3}} g_5 g_{\mu\nu},
		\nonumber \\
	H_5^+ W_{\mu}^- Z_{\nu}: &\quad & - i \frac{2 M_W M_Z}{v} \frac{g_5}{\sqrt{2}} g_{\mu\nu},
		\nonumber \\
	H_5^{++} W_{\mu}^- W_{\nu}^-: &\quad & i \frac{2 M_W^2}{v} g_5 g_{\mu\nu},
	\label{eq:H5VV}
\end{eqnarray}
where the coupling strength $g_5$ will be given in terms of $s_H$ for each model in Eq.~(\ref{eq:g5}).
These couplings imply a simple relationship among the $H_5$ decay widths to vector bosons in the high-mass limit $m_5 \gg M_{W,Z}$,
\begin{equation}
	\Gamma(H_5^{++} \to W^+ W^+) \simeq \Gamma(H_5^+ \to W^+Z)
	\simeq \Gamma(H_5^0 \to W^+W^- + ZZ) \simeq \frac{g_5^2 m_5^3}{32 \pi v^2},
	\label{eq:VVwidths}
\end{equation}
with $\Gamma(H_5^0 \to ZZ) \simeq 2 \Gamma(H_5^0 \to W^+W^-)$.\footnote{This last expression is in contrast to the case of a heavy SM Higgs boson, in which $\Gamma(h^{\rm SM} \to ZZ) \simeq \frac{1}{2} \Gamma(h^{\rm SM} \to W^+W^-)$.}

Unitarity of the longitudinal $VV \to VV$ amplitudes fixes $g_5$ in terms of $\kappa_V^h$ and $\kappa_V^H$~\cite{Falkowski:2012vh,Grinstein:2013fia}:
\begin{equation}
	g_5^2 = \frac{6}{5} (a^2 - 1),
	\label{eq:g51}
\end{equation}
where
\begin{equation}
	a^2 = (\kappa_V^h)^2 + (\kappa_V^H)^2 = 1 + (A - 1) s_H^2,
\end{equation}
with $A$ as given in Eq.~(\ref{eq:Avalues}).

This relation can be re-expressed as a sum rule for the couplings~\cite{HHG}:
\begin{equation}
	(\kappa_V^h)^2 + (\kappa_V^H)^2 - \frac{5}{6} (g_5)^2 = 1.
	\label{eq:g52}
\end{equation}
In the familiar two Higgs doublet model, where $g_5 \equiv 0$ because there is no custodial fiveplet, this reduces to the usual sum rule $(\kappa_V^h)^2 + (\kappa_V^H)^2 = 1$ for the two CP-even neutral Higgs boson couplings~\cite{Gunion:1990kf}.

Equation~(\ref{eq:g51}) yields the following values for $g_5$ in each of the models:
\begin{eqnarray}
	{\rm GM}: &\qquad& g_5 = \sqrt{2} \, s_H,  \nonumber \\
	{\rm GGM4}: &\qquad& g_5 = \sqrt{\frac{24}{5}} \, s_H, \nonumber \\
	{\rm GGM5}: &\qquad& g_5 = \sqrt{\frac{42}{5}} \, s_H, \nonumber \\
	{\rm GGM6}: &\qquad& g_5 = \frac{8}{\sqrt{5}} \, s_H.
	\label{eq:g5}
\end{eqnarray}
We note in particular that, even for fixed $s_H$, the coupling strength of $H_5$ to $VV$ grows with increasing size of the $(n \times n)$ representation.  This implies that the constraints on $s_H$ as a function of $m_5$ from $H_5^{++}$ production in vector boson fusion~\cite{Chiang:2014bia} will be more stringent in the generalized GM models than in the original GM model.  This will be illustrated in the next subsection.

The finite piece of the longitudinal $VV \to VV$ scattering amplitudes, which remains constant in the high-energy limit, can also be used to constrain the generalized GM models.  In the SM, this finite piece yields the famous constraint on the SM Higgs mass~\cite{Lee:1977yc}, $m_{h^{\rm SM}}^2 < 16 \pi v^2 /5$, where we include the contributions from the coupled channels $W^+W^- \to W^+W^-$, $W^+ W^- \leftrightarrow ZZ$, and $ZZ \to ZZ$ and require $|{\rm Re}\, a_0| < 1/2$.  In the generalized GM models, this unitarity constraint becomes
\begin{equation}
	\left[ (\kappa_V^h)^2 m_h^2 + (\kappa_V^H)^2 m_H^2 + \frac{2}{3} g_5^2 m_5^2 \right]
		< \frac{16 \pi v^2}{5}.
\end{equation}
Together with Eq.~(\ref{eq:g52}), this constraint can be recast as an upper bound on $\kappa_V^h$ or on $s_H$, as a function of $m_5$.  Setting $\kappa_V^H = 0$, we obtain absolute upper bounds on $\kappa_V^h$ and $g_5$ from perturbative unitarity of $VV \to VV$ scattering amplitudes,
\begin{equation}
	(\kappa_V^h)^2 < \frac{(16 \pi v^2 - 5 m_h^2)}{(4 m_5^2 + 5 m_h^2)} + 1, \qquad \qquad
	g_5^2 < \frac{6}{5} \frac{(16\pi v^2 - 5 m_h^2)}{(4 m_5^2 + 5 m_h^2)}.
	\label{eq:VVbounds}
\end{equation}
The bound on $g_5$ can be translated into a bound on $s_H$ in each model using Eq.~(\ref{eq:g5}).  It also leads to a very simple upper bound on the widths of the $H_5$ states given in Eq.~(\ref{eq:VVwidths}) for $m_5 \gg M_{W,Z}, m_h$,
\begin{equation}
	\Gamma(H_5 \to VV) \lesssim \frac{3}{20} m_5.
\end{equation}

The range of $\kappa_V^h$ that is actually populated in the GM model after imposing all theoretical constraints is significantly smaller than the bound from $VV \to VV$ perturbative unitarity given in Eq.~(\ref{eq:VVbounds}); for example, for $m_5 = 1000$~GeV, the maximum allowed value of $\kappa_V^h$ is about 1.1~\cite{Hartling:2014zca}, while Eq.~(\ref{eq:VVbounds}) yields an upper limit of about 1.4.  Nevertheless, in the absence of a full study of the theoretical constraints on the generalized GM models, this $VV \to VV$ unitarity bound provides a useful constraint in the high $m_5$ region that is nicely complementary to the direct constraints from $H_5^{++}$ searches, as we will show in the next section.

\subsection{Experimental constraints}
\label{sec:expt}

\begin{figure}
\resizebox{0.5\textwidth}{!}{\includegraphics{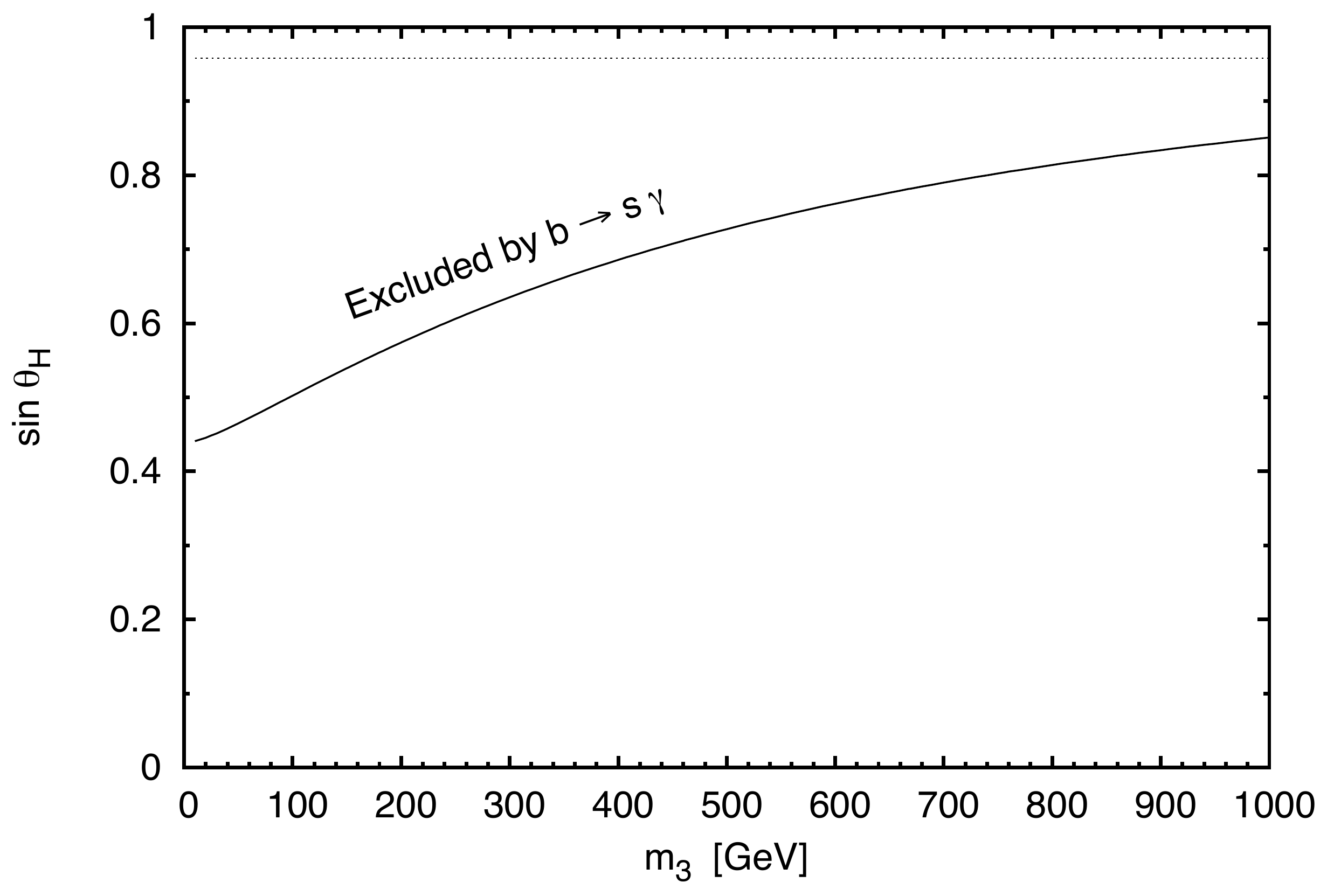}}%
\resizebox{0.5\textwidth}{!}{\includegraphics{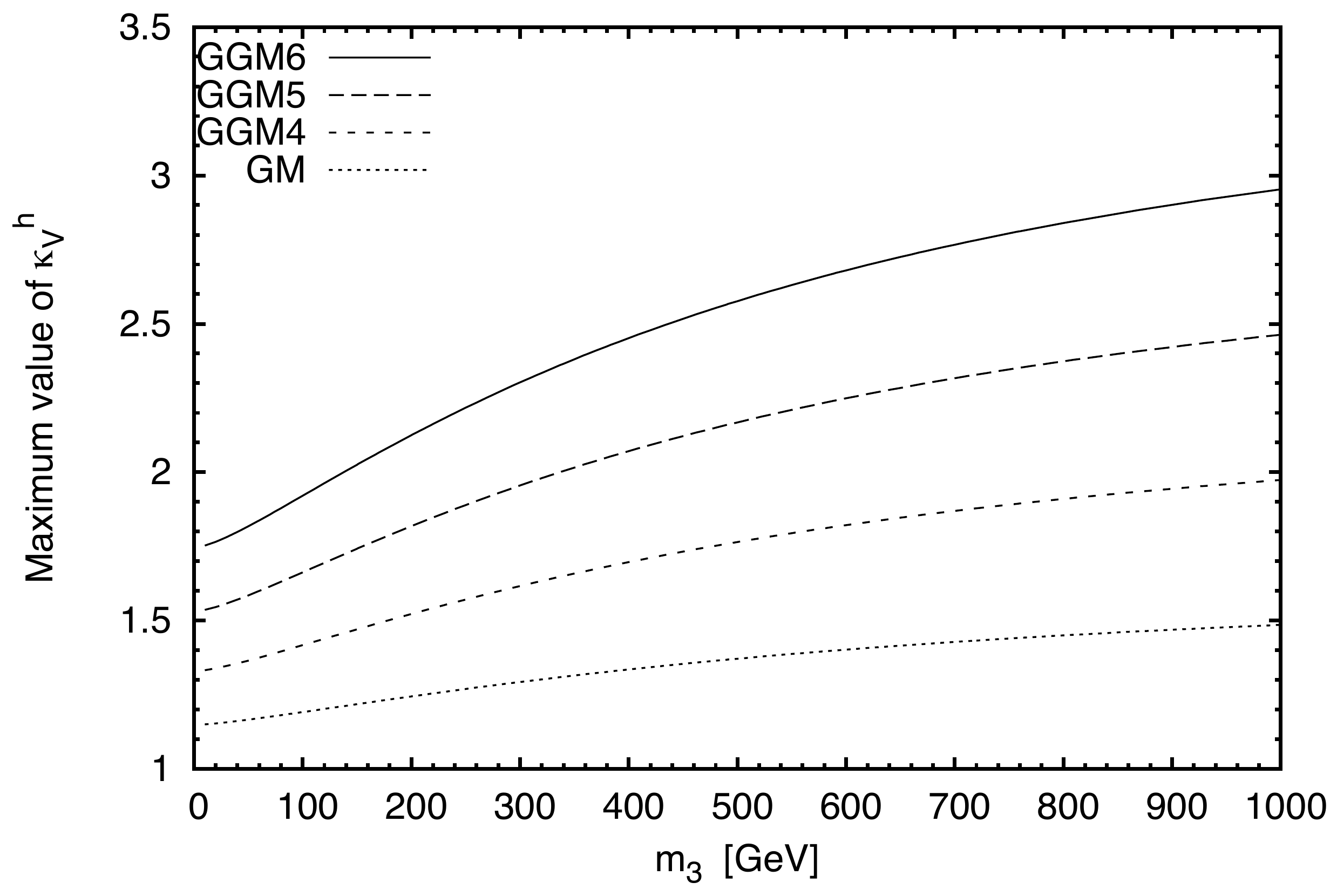}}
\caption{Left: upper bound on $s_H$ as a function of $m_3$ imposed by the ``loose'' $b \to s \gamma$ constraint of Ref.~\cite{Hartling:2014aga}.  The same bound applies to all the generalized GM models.  The horizontal dotted line near the top of this plot indicates the constraint $\tan\theta_H < 10/3$~\cite{Barger:1989fj}.  Right: maximum value of $\kappa_V^h$ as a function of $m_3$ in the GM model and the three generalized GM models, after imposing the $b \to s \gamma$ constraint.}
\label{fig:bsg}
\end{figure}

Experimental constraints on the $H_3^+$ mass and Yukawa couplings from $b \to s \gamma$ were studied in the GM model in Ref.~\cite{Hartling:2014aga}.  Re-expressing the conservative ``loose'' bound from $b \to s \gamma$ from Ref.~\cite{Hartling:2014aga} in terms of $s_H$ yields an upper bound on $s_H$ as a function of $m_3$ as shown in the left panel of Fig.~\ref{fig:bsg}.  Even for $H_3^+$ masses as high as 1~TeV, the constraint from $b \to s \gamma$ is still considerably more restrictive than the limit $\tan\theta_H < 10/3$~\cite{Barger:1989fj} imposed to avoid nonperturbative values of the top quark Yukawa coupling, which is shown by the horizontal dotted line near the top of the left panel of Fig.~\ref{fig:bsg}.

The effect of the $b \to s \gamma$ constraint on the maximum value of $\kappa_V^h$ is shown in the right panel of Fig.~\ref{fig:bsg}.  By restricting $s_H$, the $b \to s \gamma$ constraint reduces the maximum possible value of $\kappa_V^h$ compared to the values in Eq.~(\ref{eq:kappabounds}).

\begin{figure}
\resizebox{0.5\textwidth}{!}{\includegraphics{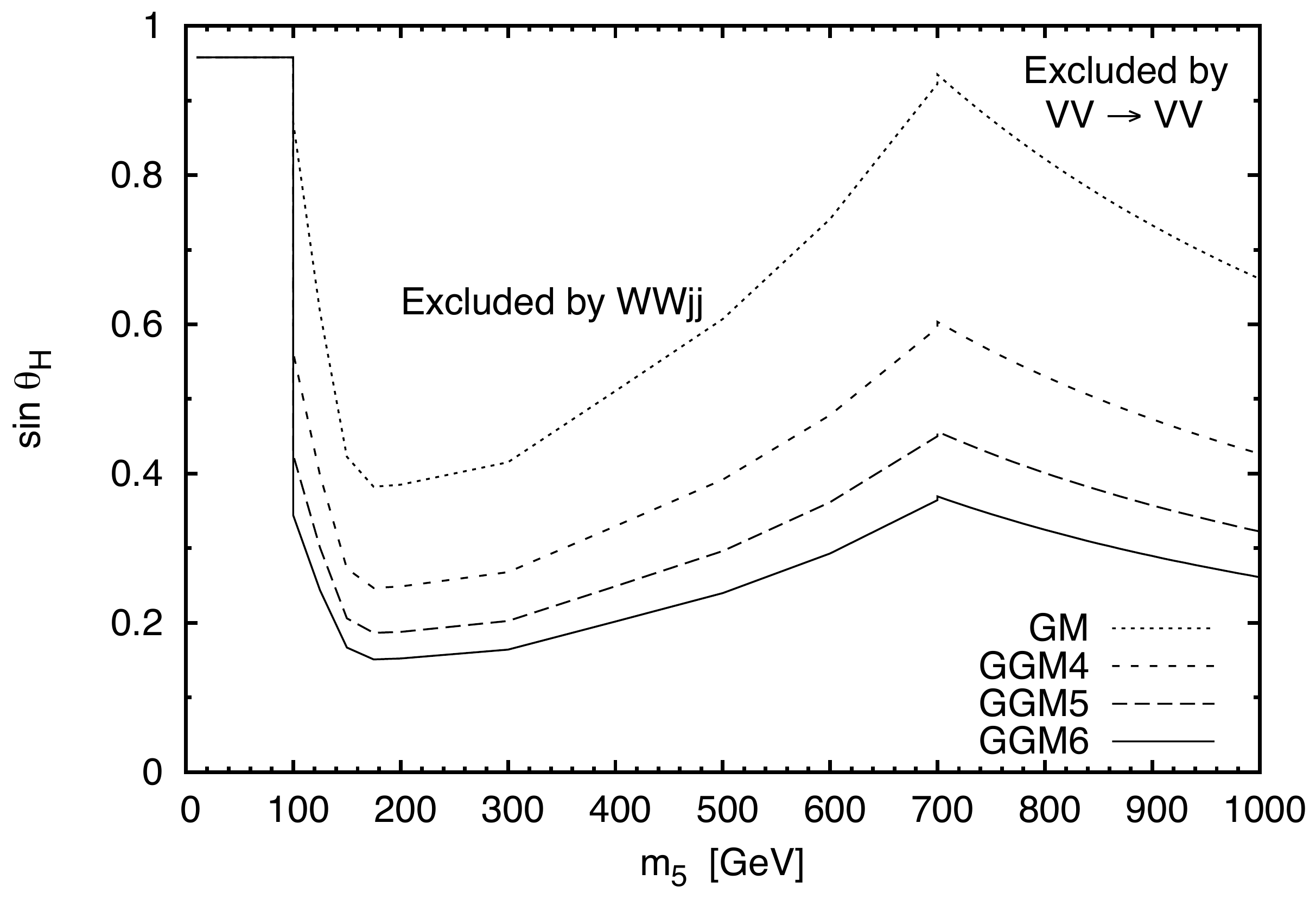}}%
\resizebox{0.5\textwidth}{!}{\includegraphics{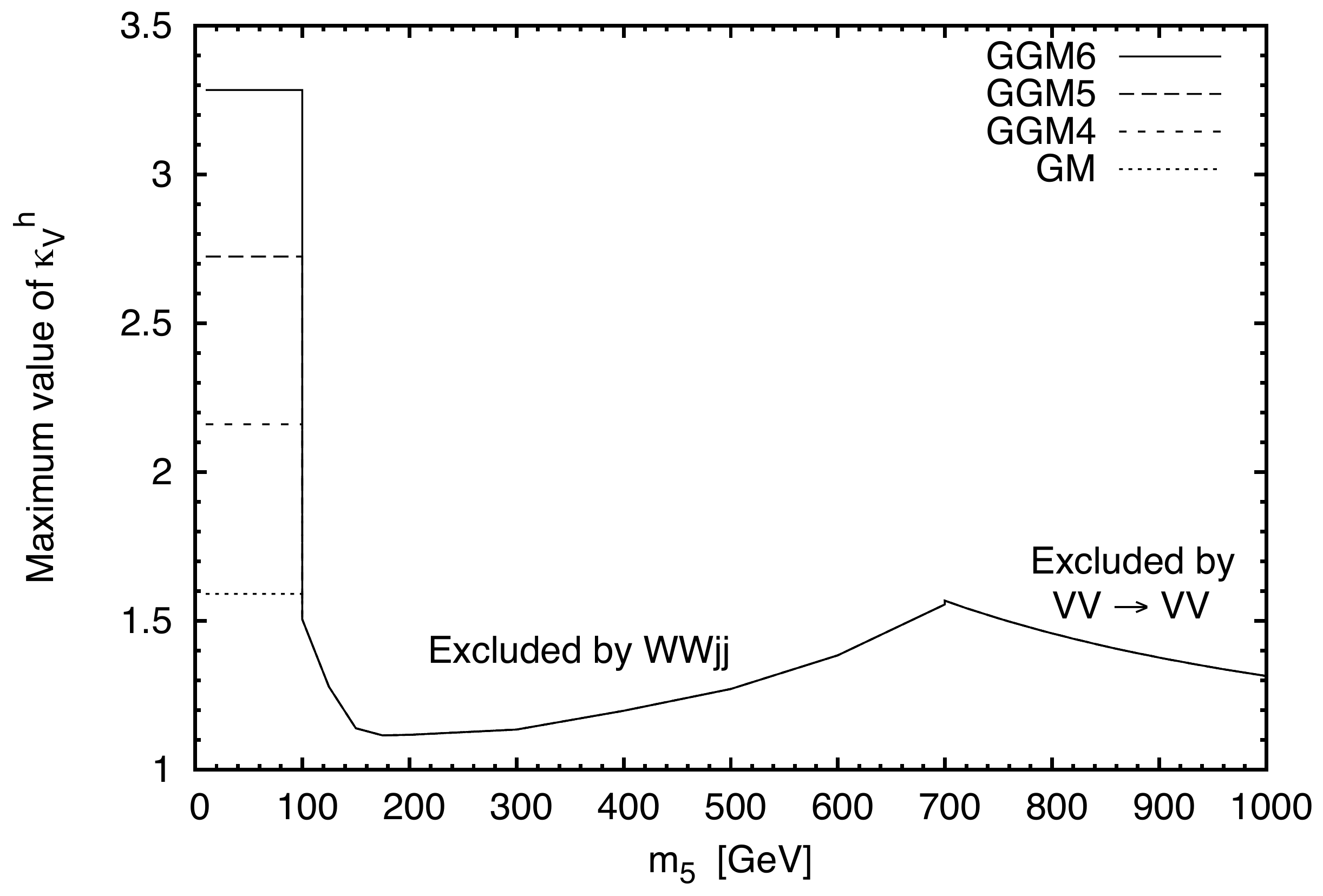}}
\caption{Constraints on the GM model and the three generalized GM models for $m_5 \geq 100$~GeV (constraints for $m_5 < 100$~GeV are shown in Fig.~\ref{fig:low}).  Left: upper bound on $s_H$ as a function of $m_5$ imposed by the like-sign $WWjj$ cross section constraint of Ref.~\cite{Chiang:2014bia}, assuming ${\rm BR}(H_5^{++} \to W^+ W^+) = 1$.  The upper bound shown for $m_5 < 100$~GeV is the constraint $\tan\theta_H < 10/3$~\cite{Barger:1989fj}.  For $m_5 > 700$~GeV, the strongest constraint comes from requiring perturbative unitarity of the finite part of the $VV \to VV$ scattering amplitude.  Right: maximum value of $\kappa_V^h$ as a function of $m_5$, after imposing the like-sign $WWjj$ cross section constraint from Ref.~\cite{Chiang:2014bia} and the $VV\to VV$ unitarity constraint from Eq.~(\ref{eq:VVbounds}).}
\label{fig:wwjj}
\end{figure}

Experimental constraints on the $H_5^{++}$ mass and its coupling to $W^+W^+$ were studied in the GM model in Ref.~\cite{Chiang:2014bia} by recasting an ATLAS measurement of the like-sign $WWjj$ cross section.  The limit in Ref.~\cite{Chiang:2014bia} assumes that ${\rm BR}(H_5^{++} \to W^+ W^+) = 1$, which can be ensured by making $m_3$, $m_7$, etc.\ larger than $m_5$.  Re-expressing the bound of Ref.~\cite{Chiang:2014bia} in terms of $g_5$ renders it independent of the size of the $(n \times n)$ representation, because the cross section depends only on the $H_5^{++} W^- W^-$ coupling as given in Eq.~(\ref{eq:H5VV}).  This bound on $g_5$ can then be translated into upper bounds on $s_H$ in each model using Eq.~(\ref{eq:g5}).  Results are shown in the left panel of Fig.~\ref{fig:wwjj}.

The effect of the like-sign $WWjj$ cross section measurement on the maximum value of $\kappa_V^h$ is shown in the right panel of Fig.~\ref{fig:wwjj}.  In particular, this constraint is the same in all the generalized GM models, independent of the size of the $(n \times n)$ representation.  This is because Eq.~(\ref{eq:g52}) directly relates the maximum allowed value of $g_5$ to the maximum allowed value of $\kappa_V^h$, independent of the size of the $(n \times n)$ representation.  The measurement provides a quite stringent constraint on $\kappa_V^h$ for $m_5$ values between 100~and 700~GeV.

In Fig.~\ref{fig:wwjj} we also show the constraints on $s_H$ and the maximum value of $\kappa_V^h$ from requiring perturbative unitarity of the finite part of the $VV \to VV$ scattering amplitude, as given by Eq.~(\ref{eq:VVbounds}).  This provides the strongest constraint on the models for $m_5$ above 700~GeV.

\begin{figure}
\resizebox{0.5\textwidth}{!}{\includegraphics{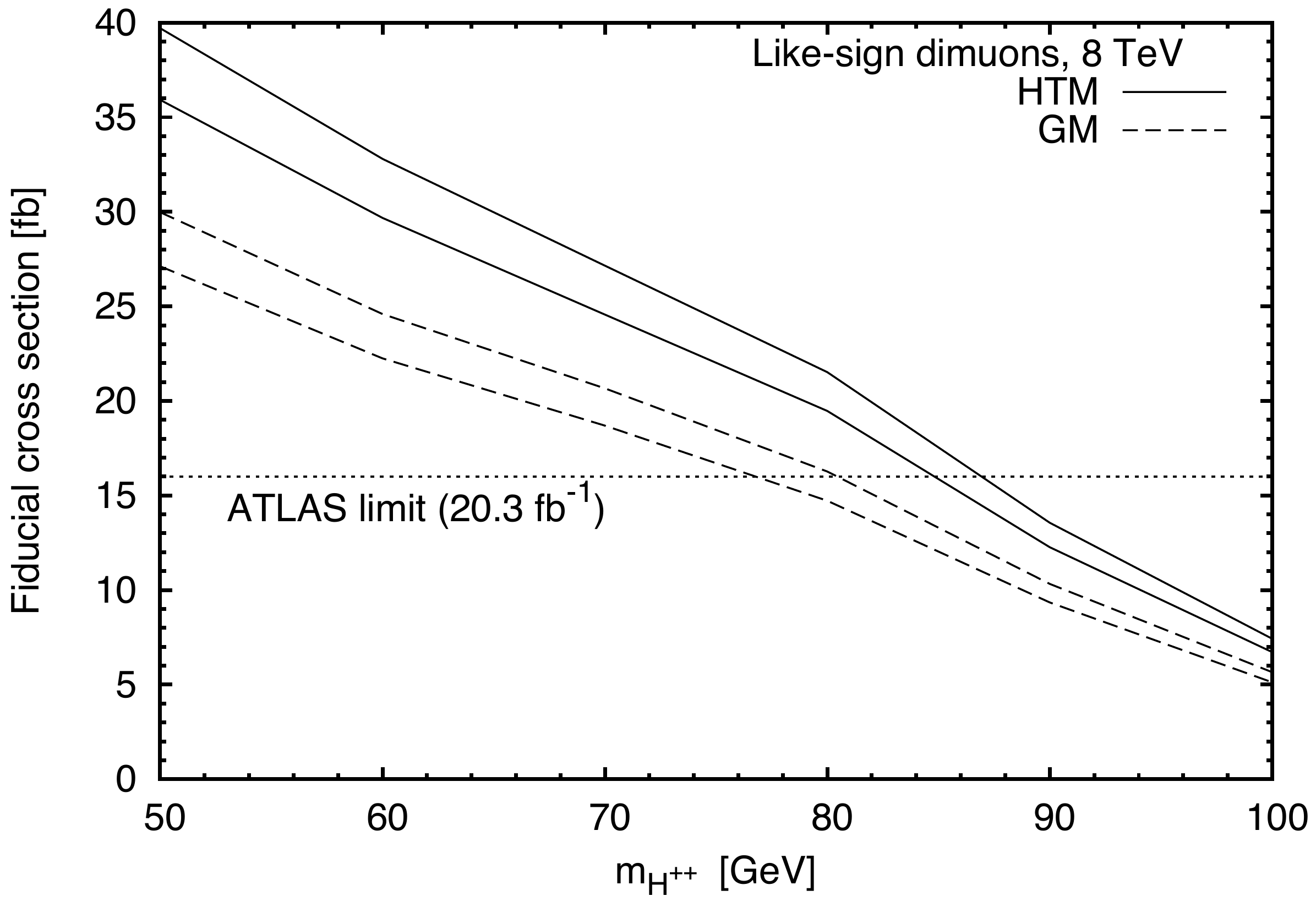}}
\caption{Fiducial cross section for the $\mu^{\pm}\mu^{\pm}$ final state from $H_5^{++}H_5^{--}$ and $H_5^{\pm\pm}H_5^{\mp}$ pair production at the 8~TeV LHC as a function of $m_{H^{++}} = m_5$, assuming BR($H_5^{++} \to W^+ W^+) = 1$, as adapted from the results of Ref.~\cite{Kanemura:2014ipa} for the HTM.  The horizontal dotted line shows the 95\% confidence level upper limit from ATLAS with 20.3~fb$^{-1}$ of data~\cite{ATLAS:2014kca}.  The widths of the bands represent a $\pm 5\%$ theoretical uncertainty on the cross sections.  This yields $m_5 \gtrsim 76$~GeV in the GM model and its generalizations, independent of the value of $s_H$.}
\label{fig:htm}
\end{figure}

Finally, we apply two further constraints that rely on the presence of $H_5^+$ and $H_5^0$ degenerate in mass with $H_5^{++}$.  First, an absolute lower bound on the doubly-charged scalar mass from ATLAS like-sign dimuon data was recently obtained in Ref.~\cite{Kanemura:2014ipa} for the Higgs Triplet Model (HTM)~\cite{HTM}, in which the SM is extended by a single complex isospin-triplet scalar field with $Y=2$, assuming that BR($H^{++} \to W^+ W^+) = 1$ and that the singly-charged scalar has the same mass as the doubly-charged scalar.  In the GM model and its generalizations, the relevant production cross sections, evaluated at next-to-leading order (NLO) in QCD, are rescaled compared to those in the HTM according to\footnote{The relevant Feynman rules in the GM model and its generalizations are fixed by custodial symmetry to be
\begin{equation}
	Z_{\mu} H_5^{++} H_5^{--}: \quad i \frac{e}{s_Wc_W} (1 - 2 s_W^2) (p_1 - p_2)_{\mu},
	\qquad \qquad \qquad
	W_{\mu}^+ H_5^+ H_5^{--}: \quad i \frac{g}{\sqrt{2}} (p_1 - p_2)_{\mu},
\end{equation}
where $p_1$ and $p_2$ are the incoming momenta of the first and second scalars listed.
For comparison, the corresponding Feynman rules in the HTM are
\begin{equation}
	Z_{\mu} H^{++} H^{--}: \quad i \frac{e}{s_Wc_W} (1 - 2 s_W^2) (p_1 - p_2)_{\mu},
	\qquad \qquad \qquad
	W_{\mu}^+ H^+ H^{--}: \quad i g (p_1 - p_2)_{\mu}.
\end{equation}}
\begin{eqnarray}
	\sigma_{\rm tot}^{\rm NLO}(pp \to H_5^{++} H_5^{--})_{\rm GM}
		&=& \sigma_{\rm tot}^{\rm NLO}(pp \to H^{++} H^{--})_{\rm HTM}, \nonumber \\
	\sigma_{\rm tot}^{\rm NLO}(pp \to H_5^{\pm\pm} H_5^{\mp})_{\rm GM}
		&=& \frac{1}{2} \sigma_{\rm tot}^{\rm NLO}(pp \to H^{\pm\pm} H^{\mp})_{\rm HTM}.
\end{eqnarray}
We ignore the cross section contributions from associated production of $H_5^{\pm\pm}H_3^{\mp}$ or $H_5^{\pm\pm} H_7^{\mp}$, as well as single production of $H_5^{\pm\pm}$.  Rescaling the HTM total cross sections and reassembling the fiducial cross section from the information provided in Table~I of Ref.~\cite{Kanemura:2014ipa} yields the results shown in Fig.~\ref{fig:htm}, where the widths of the two bands represent $\pm 5\%$ theoretical uncertainty from QCD and parton distribution functions~\cite{Kanemura:2014ipa}.  Because of the reduced cross section in the GM model and its generalizations, the $H^{++}$ mass lower bound of 84~GeV found for the HTM in Ref.~\cite{Kanemura:2014ipa} is weakened to $m_5 \gtrsim 76$~GeV in the GM model and its generalizations.

\begin{figure}
\resizebox{0.5\textwidth}{!}{\includegraphics{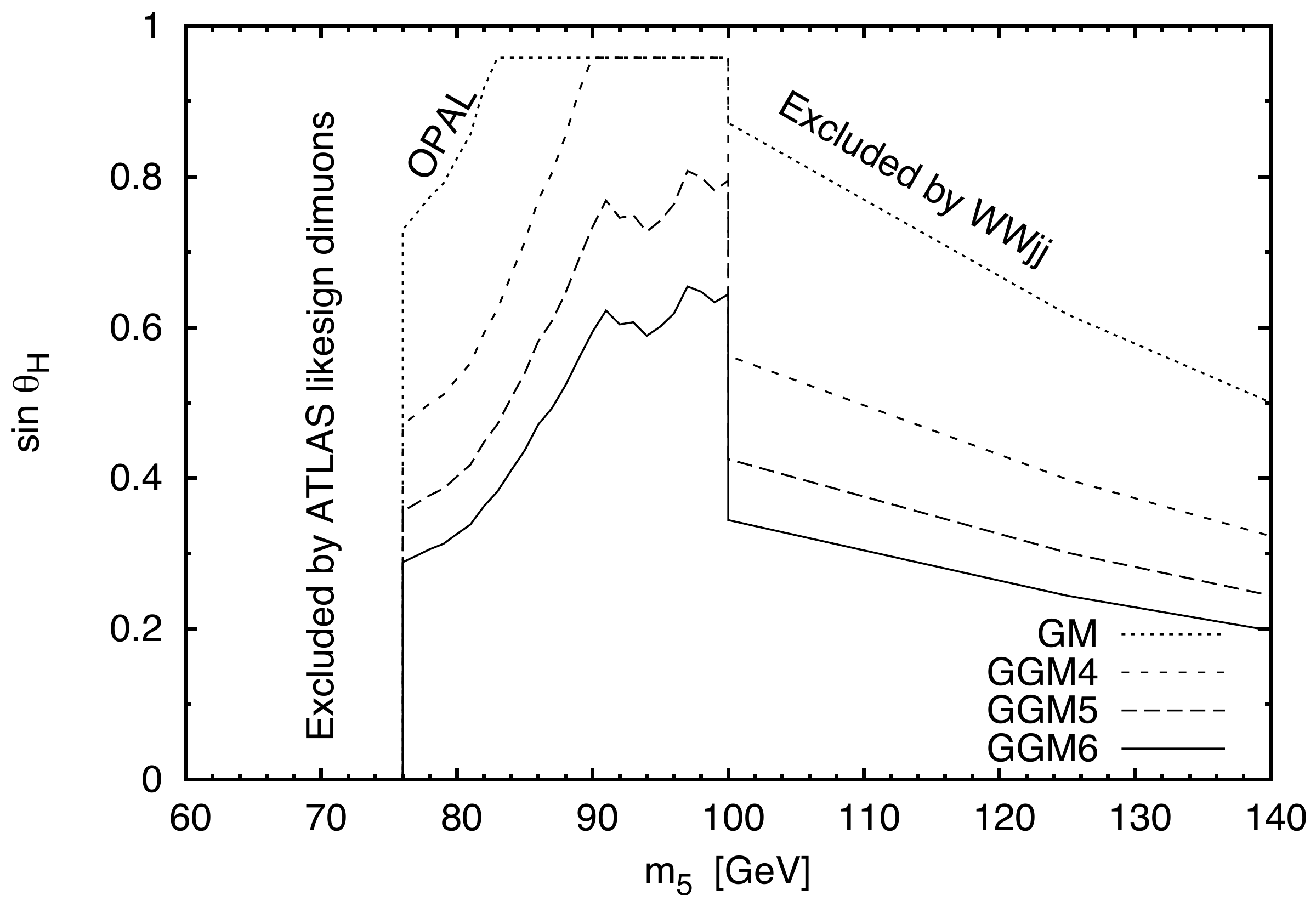}}%
\resizebox{0.5\textwidth}{!}{\includegraphics{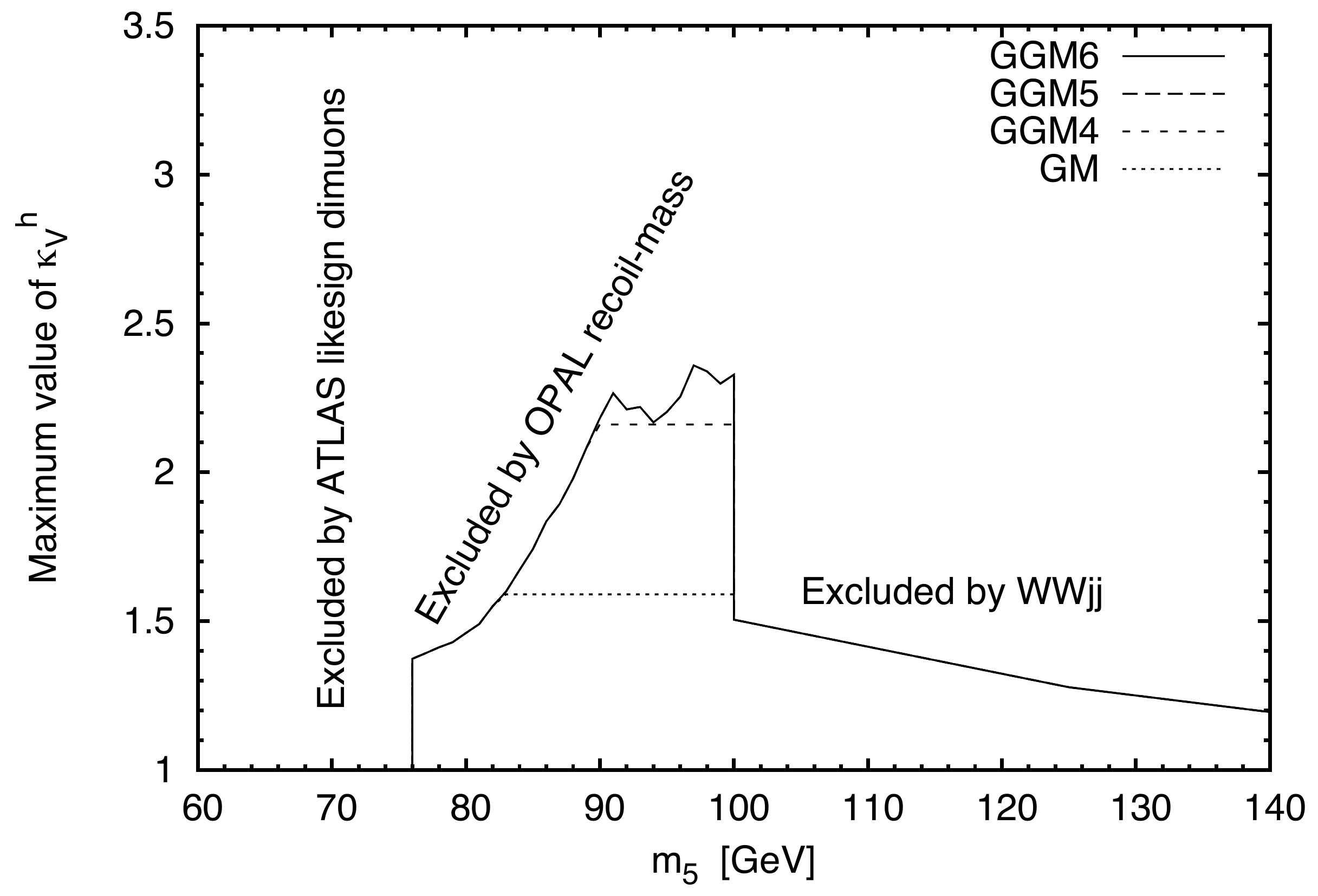}}
\caption{Constraints on the GM model and the three generalized GM models, focusing on low values of $m_5$.  Masses below about 76~GeV are excluded by ATLAS like-sign dimuon data assuming BR$(H_5^{++} \to W^+ W^+) = 1$, as shown in Fig.~\ref{fig:htm}.  For $m_5$ between 76~and 100~GeV, the strongest constraint comes from the OPAL decay-mode-independent search for $Z H_5^0$ production~\cite{Abbiendi:2002qp}; where this constraint is weak we impose $\tan\theta_H < 10/3$~\cite{Barger:1989fj}.  For $m_5$ above 100~GeV we show the like-sign $WWjj$ cross section constraint from Ref.~\cite{Chiang:2014bia} as in Fig.~\ref{fig:wwjj}.  Left: upper bound on $s_H$ as a function of $m_5$.  Right: maximum value of $\kappa_V^h$ as a function of $m_5$.}
\label{fig:low}
\end{figure}

Second, a nontrivial upper bound on $s_H$ for $m_5 \leq 100$~GeV can be obtained using the results of a decay-mode-independent search for new scalar bosons produced in association with a $Z$ boson~\cite{Abbiendi:2002qp} from OPAL at the CERN Large Electron-Positron (LEP) Collider.  This search used the recoil-mass method to set a limit on the production cross section of new scalar resonances without any reference to the decay modes of the scalar. We used the numerical tabulation of the OPAL limit in the data file {\tt lep\_decaymodeindep.txt} provided with the public code HiggsBounds 4.2.0~\cite{HiggsBounds} to constrain the $H_5^0 ZZ$ coupling [Eq.~(\ref{eq:H5VV})] as a function of $m_5$.  Results are shown in Fig.~\ref{fig:low}.  The OPAL measurement limits the maximum possible value of $\kappa_V^h$ in the GM model and its three generalizations to 2.36, which is obtained in the GGM5 and GGM6 models for $m_5 \simeq 97$~GeV.

We emphasize that these constraints rely on the presence of the custodial SU(2) symmetry in the scalar potential.  They are thus valid in the GM model and its generalizations, but they do not apply in, e.g., the septet model.

\section{Scalar potentials and decoupling behavior}
\label{sec:potentials}

We now proceed to write down the most general SU(2)$_L \times$SU(2)$_R$-invariant scalar potentials for the GGM4, GGM5, and GGM6 models.  We denote the bidoublet as $\Phi$ as in the original GM model, and the $(n \times n)$ representation with isospins $T = (n-1)/2$ as $X_n$.  (Where it will be unambiguous, we suppress the subscript on $X$ for compactness.)  We compute the minimization conditions and the physical masses in the custodial-SU(2)-preserving phase in terms of the Lagrangian parameters.

We also briefly discuss the decoupling behavior of the models.  An extension of the SM is said to possess a \emph{decoupling limit} if all the new particles can be taken arbitrarily heavy while all amplitudes involving the remaining light particles in the initial and final state approach their SM values~\cite{Appelquist:1974tg,HaberDecoupling}.  In particular, this implies that all the new particles can be taken heavy without any couplings becoming nonperturbatively large.  It also generically implies that the couplings of the remaining light SM-like Higgs boson to other SM particles will deviate from their SM values by a relative correction of order $(v/M_{\rm new})^k$, where $M_{\rm new}$ is the mass scale of the new heavy particles.  For a perturbative theory, this is equivalent to the statement that there exists an effective theory below the scale of the heavy new particles containing only the SM field content; SM gauge invariance then requires $k \geq 2$~\cite{Buchmuller:1985jz,Grzadkowski:2010es}.

The original GM model is known to possess a decoupling limit~\cite{Hartling:2014zca}.  We will show that the GGM4 model similarly possesses a decoupling limit, and highlight some differences in the rate of decoupling (equivalently the order at which the Higgs coupling modifications appear in the effective field theory) compared to the GM model.  We will also show that the electroweak symmetry breaking vacuum in the GGM5 and GGM6 models possesses two phases, one in which $v_n \neq 0$ and one in which $v_n = 0$.  In the $v_n \neq 0$ phase these two models do not possess a decoupling limit: the masses of the additional scalars are bounded from above when the scalar quartic couplings are kept perturbative.  In the $v_n = 0$ phase these two models do possess a decoupling limit in which all the additional scalars can be taken heavy while keeping all couplings perturbative.

\subsection{GGM4}
\label{sec:ggm4}

The scalar potential for the GGM4 model can be written as (repeated indices are always summed)
\begin{eqnarray}
	V(\Phi, X) &=& \frac{m^2_\Phi}{2} {\rm Tr} ( \Phi^\dagger \Phi )
	+ \frac{m^2_X}{2} {\rm Tr}( X^\dagger X)
	+  \lambda_1 \left[{\rm Tr}( \Phi^\dagger \Phi) \right]^2 \nonumber \\
	&& + \lambda_2 \left[{\rm Tr}( X^\dagger X) \right]^2
	+ \lambda_3 {\rm Tr}( X^\dagger X X^\dagger X)
	+  \lambda_4 {\rm Tr}( X^\dagger T_{3/2}^a X T_{3/2}^b){\rm Tr}( X^\dagger T_{3/2}^a X T_{3/2}^b) \nonumber\\
	&& + \lambda_5 {\rm Tr}( \Phi^\dagger \Phi){\rm Tr}( X^\dagger X) + \lambda_6 {\rm Tr}( \Phi^\dagger T_{1/2}^a \Phi T_{1/2}^b){\rm Tr}( X^\dagger T_{3/2}^a X T_{3/2}^b) \nonumber \\
	&& + \lambda_7 {\rm Tr}\left[ \Phi^\dagger \widehat{T}^{1,i}_{1/2} \Phi (\widehat{T}^{1,j}_{1/2})^\dagger\right] {\rm Tr}\left[ \Phi^\dagger (\widehat{T}^{1,i}_{3/2,1/2})^\dagger X  \widehat{T}^{1,j}_{3/2,1/2}\right] \nonumber \\
	&& + \lambda_8 {\rm Tr}\left[ X^\dagger \widehat{T}^{1,i}_{3/2} X  (\widehat{T}^{1,j}_{3/2})^\dagger \right] {\rm Tr}\left[ \Phi^\dagger (\widehat{T}^{1,i}_{3/2,1/2})^\dagger X  \widehat{T}^{1,j}_{3/2,1/2}\right].
	\label{eq:GGM4pot}
\end{eqnarray}
The first line of this expression contains the two mass-squared terms and the doublet quartic coupling, just as in the GM model.  The second line contains the three linearly independent terms involving four powers of the $X_4$ field (in the GM model there are only two such terms).  The third line contains the two $\Phi^2 X^2$ terms; there are always only two ways to construct such terms, since the two doublets can be combined with total isospin zero or one.
The last two lines contain the terms that break the would-be $Z_2$ symmetry under which $X_4 \to -X_4$: one of the form $\Phi^3 X$ and the other of the form $\Phi X^3$.  These two terms are written in terms of the spherical tensors $\widehat T$ defined in Appendix~\ref{app:spherical}.

Minimizing the potential while assuming that custodial SU(2) is not spontaneously broken gives the two constraint equations,
\begin{eqnarray}
	0 = \frac{\partial V}{\partial v_{\phi}} &=&
		m_{\Phi}^2 v_{\phi} + 4 \lambda_1 v_{\phi}^3
		+ \frac{1}{2} \left[ 16 \lambda_5 + 15 \lambda_6 \right] v_{\phi} v_4^2
		+ \frac{9}{4 \sqrt{2}} \lambda_7 v_{\phi}^2 v_4
		+ \frac{15}{\sqrt{2}} \lambda_8 v_4^3, \nonumber \\
	0 = \frac{\partial V}{\partial v_4} &=&
		4 m_X^2 v_4 + 4 \left[ 16 \lambda_2 + 4 \lambda_3 + 75 \lambda_4 \right] v_4^3
		+ \frac{1}{2} \left[ 16 \lambda_5 + 15 \lambda_6 \right] v_{\phi}^2 v_4
		+ \frac{3}{4 \sqrt{2}} \lambda_7 v_{\phi}^3 + \frac{45}{\sqrt{2}} \lambda_8 v_{\phi} v_4^2.
\end{eqnarray}
When both $\lambda_7$ and $\lambda_8$ are nonzero, there are only two phases: one in which $v_{\phi} = v_4 = 0$ and one in which both $v_{\phi}$ and $v_4$ are nonzero.  In the latter phase we can solve for $m^2_{\Phi}$ and $m^2_X$,
\begin{eqnarray}
	m^2_{\Phi} &=& -4 \lambda_1 v_{\phi}^2
	- \frac{1}{2} \left[ 16 \lambda_5 + 15 \lambda_6 \right] v_4^2
	- \frac{9}{4 \sqrt{2}} \lambda_7 v_{\phi} v_4
	- \frac{15}{\sqrt{2}} \lambda_8 \frac{v_4^3}{v_{\phi}}, \nonumber \\
	m^2_X &=& - \left[16 \lambda_2 + 4 \lambda_3 + 75 \lambda_4 \right] v_4^2
	- \frac{1}{8} \left[ 16 \lambda_5 + 15 \lambda_6 \right] v_{\phi}^2
	- \frac{3}{16 \sqrt{2}} \lambda_7 \frac{v_{\phi}^3}{v_4}
	- \frac{45}{4 \sqrt{2}} \lambda_8 v_{\phi} v_4.
	\label{eq:GGM4mincond}
\end{eqnarray}
The vevs are further constrained by the $W$ mass to obey
\begin{equation}
	v_{\phi}^2 + 20 v_4^2 = v^2,
\end{equation}
where $v^2$ is the SM Higgs vev given in Eq.~(\ref{eq:vevrelation}).

Using Eqs.~(\ref{eq:GGM4mincond}) to eliminate $m_{\Phi}^2$ and $m_X^2$, the masses of the physical states in the custodial sevenplet, fiveplet, and triplet are given by
\begin{eqnarray}
	m_7^2 &=& -120 \lambda_4 v_4^2 - 3 \lambda_6 v_{\phi}^2
		- \frac{3}{16 \sqrt{2}} \lambda_7 \frac{v_{\phi}^3}{v_4}
		- \frac{63}{4 \sqrt{2}} \lambda_8 v_{\phi} v_4, \nonumber \\
	m_5^2 &=& 4 \left[2 \lambda_3 - 3 \lambda_4 \right] v_4^2
		- \frac{3}{2} \lambda_6 v_{\phi}^2
		- \frac{3}{16 \sqrt{2}} \lambda_7 \frac{v_{\phi}^3}{v_4}
		- \frac{63}{4 \sqrt{2}} \lambda_8 v_{\phi} v_4, \nonumber \\
	m_3^2 &=& - (v_{\phi}^2 + 20 v_4^2) \left[ \frac{1}{2} \lambda_6
		+ \frac{3}{16 \sqrt{2}} \lambda_7 \frac{v_{\phi}}{v_4}
		+ \frac{3}{4 \sqrt{2}} \lambda_8 \frac{v_4}{v_{\phi}} \right].
\end{eqnarray}
The elements of the custodial-singlet mass-squared matrix in the $(\phi^{0,r}, H_1^{\prime 0})$ basis are given by,
\begin{eqnarray}
	\mathcal{M}_{11}^2 &=& 8 \lambda_1 v_{\phi}^2
		+ \frac{9}{4 \sqrt{2}} \lambda_7 v_{\phi} v_4
		- \frac{15}{\sqrt{2}} \lambda_8 \frac{v_4^3}{v_{\phi}}, \nonumber \\
	\mathcal{M}_{12}^2 &=& \frac{1}{2} \left[ 16 \lambda_5 + 15 \lambda_6 \right] v_{\phi} v_4
		+ \frac{9}{8 \sqrt{2}} \lambda_7 v_{\phi}^2
		+ \frac{45}{2 \sqrt{2}} \lambda_8 v_4^2, \nonumber \\
	\mathcal{M}_{22}^2 &=& 2 \left[ 16 \lambda_2 + 4 \lambda_3 + 75 \lambda_4 \right] v_4^2
		- \frac{3}{16 \sqrt{2}} \lambda_7 \frac{v_{\phi}^3}{v_4}
		+ \frac{45}{4 \sqrt{2}} \lambda_8 v_{\phi} v_4.
\end{eqnarray}
The mass eigenstates and mixing angle are defined as in Eq.~(\ref{eq:hmass}).
The compositions of the physical states are given explicitly in Appendix~\ref{app:states4}.

The GGM4 model possesses a decoupling limit.  Consider the situation in which $m_{\Phi}^2 < 0$ (to break electroweak symmetry) and $m_X^2 \gg v^2$.  The $\lambda_7 \Phi^3 X$ term in Eq.~(\ref{eq:GGM4pot}) induces a small vev for $X_4$ once $\Phi$ gets its vev, $v_4 \ll v_{\phi} \simeq v$ (in fact, this term ensures $v_4 \neq 0$ unless $\lambda_7 = 0$).  The expression for $m_X^2$ in Eq.~(\ref{eq:GGM4mincond}) then implies that
\begin{equation}
	m_X^2 \simeq - \frac{3}{16 \sqrt{2}} \lambda_7 \frac{v^3}{v_4},
\end{equation}
or
\begin{equation}
	s_H = \sqrt{20} \frac{v_4}{v} \simeq - \frac{3 \sqrt{5}}{8 \sqrt{2}} \lambda_7 \frac{v^2}{m_X^2},
	\label{eq:sHdecoup}
\end{equation}
implying that the Goldstone bosons consist increasingly of isospin doublet as $m_X^2$ is taken large.

In this limit, the masses of the custodial sevenplet, fiveplet and triplet become
\begin{equation}
	m_7^2 \simeq m_5^2 \simeq m_3^2 \simeq - \frac{3}{16 \sqrt{2}} \lambda_7 \frac{v^3}{v_4} \simeq m_X^2,
\end{equation}
while the elements of the custodial-singlet mass-squared matrix become
\begin{equation}
	\mathcal{M}_{11}^2 \simeq 8 \lambda_1 v^2,  \qquad
	\mathcal{M}_{12}^2 \simeq \frac{9}{8 \sqrt{2}} \lambda_7 v^2, \qquad
	\mathcal{M}_{22}^2 \simeq - \frac{3}{16 \sqrt{2}} \lambda_7 \frac{v^3}{v_4} \simeq m_X^2.
\end{equation}
Thus the exotic scalars all become heavy with a common mass $m_X$, leaving one light state with mass $m_h \simeq 8 \lambda_1 v^2$.  The mixing angle between the two custodial-singlet states becomes small in this limit,
\begin{equation}
	s_{\alpha} \simeq \frac{9}{8 \sqrt{2}} \lambda_7 \frac{v^2}{m_X^2}
	\simeq - \frac{3}{\sqrt{5}} s_H,
\end{equation}
implying that $h$ consists increasingly of isospin doublet as $m_X^2$ is taken large.

The couplings of $h$ in this limit approach those of the SM:
\begin{equation}
	\kappa_V^h \simeq 1 + \frac{8}{5} s_H^2 \equiv 1 + 4 \epsilon, \qquad
	\kappa_f^h \simeq 1 - \frac{2}{5} s_H^2 \equiv 1 - \epsilon,
\end{equation}
where
\begin{equation}
	\epsilon \equiv \frac{2}{5} s_H^2 \simeq \frac{9}{64} \lambda_7^2 \frac{v^4}{m_X^4}.
\end{equation}
From this we can draw a number of conclusions that hold in the decoupling limit of the GGM4 model:
\begin{itemize}
\item[$(i)$] The coupling of $h$ to vector boson pairs is enhanced and its coupling to fermion pairs is suppressed compared to the SM in the decoupling limit.  This is the same pattern as in the GM model~\cite{Hartling:2014zca}.
\item[$(ii)$] The deviation of $\kappa_V^h$ from one is four times as large as that of $\kappa_f^h$ in the decoupling limit.  This differs from the original GM model, in which the deviation of $\kappa_V^h$ from one is three times as large as that of $\kappa_f^h$ in the decoupling limit~\cite{Hartling:2014zca}.
\item[$(iii)$] The size of the deviations of $\kappa_V^h$ and $\kappa_f^h$ from their SM values decouples like $v^4/m_X^4$.  This decoupling is much faster than the bound from unitarity of the finite part of the $VV \to VV$ amplitude in Eq.~(\ref{eq:VVbounds}), which requires that the deviation of $\kappa_V^h$ from its SM value decouple like $v^2/m_X^2$.  For comparison, in the two Higgs doublet model, the deviation in $\kappa_V^h$ similarly decouples like $v^4/M_A^4$, while the deviations in $\kappa_f^h$ for the various fermion species decouple much more slowly, like $v^2/M_A^2$, where $M_A$ is the mass scale of the additional scalars in the model~\cite{Gunion:2002zf}.
\end{itemize}

\subsection{GGM5}
\label{sec:ggm5}

The scalar potential for the GGM5 model can be written as
\begin{eqnarray}
	V(\Phi, X) &=& \frac{m^2_\Phi}{2} {\rm Tr} ( \Phi^\dagger \Phi )
	+ \frac{m^2_X}{2} {\rm Tr}( X^\dagger X)
	+ \lambda_1 \left[ {\rm Tr}( \Phi^\dagger \Phi) \right]^2  \nonumber \\
	&& + \lambda_2 \left[ {\rm Tr}( X^\dagger X) \right]^2
	+ \lambda_3  {\rm Tr}( X^\dagger X X^\dagger X)
	+  \lambda_4 {\rm Tr} \left[ X^\dagger \widehat{T}^{1,i}_2 X (\widehat{T}^{1,j}_2)^\dagger \right] {\rm Tr} \left[ X^\dagger (\widehat{T}^{1,i}_2)^\dagger X \widehat{T}^{1,j}_2 \right] \nonumber  \\
	&& +  \lambda_5 {\rm Tr} \left[ X^\dagger  X (\widehat{T}^{2,j}_2)^\dagger \right] {\rm Tr} \left[ X^\dagger  X \widehat{T}^{2,j}_2 \right]
	+  \lambda_6 {\rm Tr} \left[ X^\dagger \widehat{T}^{2,i}_2 X \right] {\rm Tr} \left[ X^\dagger (\widehat{T}^{2,i}_2)^\dagger X \right] \nonumber \\
	&& + \lambda_7 {\rm Tr}( \Phi^\dagger \Phi){\rm Tr}( X^\dagger X) + \lambda_8 {\rm Tr}( \Phi^\dagger T_{1/2}^a \Phi T_{1/2}^b){\rm Tr}( X^\dagger T_2^a X T_2^b) \nonumber \\
	&& + M_2 {\rm Tr } \left[ X^\dagger \widehat{T}^{2,i}_{2} X (\widehat{T}^{2,j}_{2})^\dagger \right] X_{ij}.
	\label{eq:GGM5pot}
\end{eqnarray}
The first line of this expression contains the two mass-squared terms and the doublet quartic coupling, just as in the GM model.  The second and third lines contain the five linearly independent terms involving four powers of the $X_5$ field (in the GM model there are only two such terms).  The fourth line contains the two $\Phi^2 X^2$ terms; there are always only two ways to construct such terms, since the two doublets can be combined with total isospin zero or one.
The last line contains the term of the form $X^3$ that breaks the would-be $Z_2$ symmetry under which $X_5 \to -X_5$.  Again we have used the spherical tensors $\widehat T$ defined in Appendix~\ref{app:spherical} to write the potential in a compact form.

Minimizing the potential while assuming that custodial SU(2) is not spontaneously broken gives the two constraint equations,
\begin{eqnarray}
	0 = \frac{\partial V}{\partial v_{\phi}} &=&
		v_{\phi} \left\{ m_{\Phi}^2 + 4 \lambda_1 v_{\phi}^2
		+ 5 \left[ 2 \lambda_7 + 3 \lambda_8 \right] v_5^2 \right\}, \nonumber \\
	0 = \frac{\partial V}{\partial v_5} &=&
		5 v_5 \left\{ m_X^2 + 4 \left[ 5 \lambda_2 + \lambda_3 + 60 \lambda_4 \right] v_5^2
		+ \left[ 2 \lambda_7 + 3 \lambda_8 \right] v_{\phi}^2 + 63 M_2 v_5 \right\}.
\end{eqnarray}
We will require $v_{\phi} \neq 0$ in order to generate fermion masses.  Then there are two phases: $v_5 = 0$ and $v_5 \neq 0$.  We first consider the case $v_5 \neq 0$; we will discuss the case $v_5 = 0$ below.

When both $v_{\phi}$ and $v_5$ are nonzero, we can solve for $m_{\Phi}^2$ and $m_X^2$,
\begin{eqnarray}
	m^2_{\Phi} &=& -4 \lambda_1 v_{\phi}^2 - 5 \left[2 \lambda_7 + 3 \lambda_8 \right] v_5^2,
	\nonumber \\
	m^2_X &=& -4 \left[ 5 \lambda_2 + \lambda_3 + 60 \lambda_4 \right] v_5^2
	- \left[2 \lambda_7 + 3 \lambda_8 \right] v_{\phi}^2
	- 63 M_2 v_5.
	\label{eq:GGM5mincond}
\end{eqnarray}
The vevs are further constrained by the $W$ mass to obey
\begin{equation}
	v_{\phi}^2 + 40 v_5^2 = v^2,
	\label{eq:GGM5vevs}
\end{equation}
where $v^2$ is the SM Higgs vev given in Eq.~(\ref{eq:vevrelation}).

Using Eqs.~(\ref{eq:GGM5mincond}) to eliminate $m_{\Phi}^2$ and $m_X^2$, the masses of the physical states in the custodial nineplet, sevenplet, fiveplet, and triplet are given by
\begin{eqnarray}
	m_9^2 &=& 8 \left[ \lambda_3 - 50 \lambda_4 \right] v_5^2
	- 5 \lambda_8 v_{\phi}^2 - 27 M_2 v_5 , \nonumber \\
	m_7^2 &=& -240 \lambda_4 v_5^2 - 3 \lambda_8 v_{\phi}^2 - 135 M_2 v_5, \nonumber \\
	m_5^2 &=& 8 \left[ \lambda_3 + 6 \lambda_4 + 21 \lambda_5 + 21 \lambda_6 \right] v_5^2
	- \frac{3}{2} \lambda_8 v_{\phi}^2 - 90 M_2 v_5, \nonumber \\
	m_3^2 &=& -\frac{1}{2} \lambda _8 \left(v_{\phi}^2 + 40 v_5^2 \right).
\end{eqnarray}
The elements of the custodial-singlet mass-squared matrix in the $(\phi^{0,r}, H_1^{\prime 0})$ basis are given by,
\begin{eqnarray}
	\mathcal{M}_{11}^2 &=& 8 \lambda_1 v_{\phi}^2, \nonumber \\
	\mathcal{M}_{12}^2 &=& 2 \sqrt{5} \left[ 2 \lambda_7 + 3 \lambda_8 \right] v_{\phi} v_5, \nonumber \\
	\mathcal{M}_{22}^2 &=& 8 \left[ 5 \lambda_2 + \lambda_3 + 60 \lambda_4 \right] v_5^2
	+ 63 M_2 v_5.
\end{eqnarray}
The mass eigenstates and mixing angle are defined as in Eq.~(\ref{eq:hmass}).
The compositions of the physical states are given explicitly in Appendix~\ref{app:states5}.

In the $v_5 \neq 0$ phase, the GGM5 model does \emph{not} possess a decoupling limit.  The easiest way to see this is to note that $m_3^2 = -\lambda_8 v^2/2$, which is bounded from above by the perturbativity of $\lambda_8$.  In this phase the model also possesses the strange feature that as $v_5 \to 0$, the mass of the lighter custodial-singlet scalar also goes to zero.  Thus a lower bound can be set on $v_5$ from searches for a light custodial-singlet scalar.  This will in turn set a lower bound on the deviations of the couplings of the 125~GeV Higgs from their SM values.  This phase of the model could thus in principle be entirely ruled out by a combination of precision measurements of the 125~GeV Higgs boson couplings and searches for a light custodial-singlet scalar.  Similar features are observed~\cite{Englert:2013zpa} in the original GM model if a $Z_2$ symmetry is imposed on the scalar potential, thereby eliminating the term linear in $X$ from the scalar potential.

In the other phase $v_5 = 0$, the minimization condition $\partial V/\partial v_5 = 0$ holds automatically and $m_X^2$ cannot be eliminated from the potential.  In this case there is no mixing between the isospin doublet and the exotic scalars, and the additional scalars in $X_5$ decouple when $m_X^2 \gg v^2$.  In this phase the model is not interesting from the perspective of the LHC ``flat direction'' because $\kappa_V^h = \kappa_f^h = 1$; we include it here only for completeness.  The masses are now given by [we set $v_{\phi} = v$ as required by Eq.~(\ref{eq:GGM5vevs})]
\begin{eqnarray}
	m_9^2 &=& m_X^2 + 2 \lambda_7 v^2 - 2 \lambda_8 v^2, \nonumber \\
	m_7^2 &=& m_X^2 + 2 \lambda_7 v^2, \nonumber \\
	m_5^2 &=& m_X^2 + 2 \lambda_7 v^2 + \frac{3}{2} \lambda_8 v^2, \nonumber \\
	m_3^2 &=& m_X^2 + 2 \lambda_7 v^2 + \frac{5}{2} \lambda_8 v^2, \nonumber \\
	m_H^2 &=& m_X^2 + 2 \lambda_7 v^2 + 3 \lambda_8 v^2, \nonumber \\
	m_h^2 &=& 8 \lambda_1 v^2,
\end{eqnarray}
where $h = \phi^{0,r}$ and $H = H_1^{\prime 0}$.  Note that the ordering of the masses of the exotic scalars is monotonic in their custodial SU(2) quantum number.  In this phase the model possesses a decoupling limit: when $m_X^2 \gg v^2$, all the new states become heavy with masses of order $\sqrt{m_X^2}$ while $h$ remains at the weak scale.  The couplings of $h$ to SM particles are modified only through loops involving the new scalars, the effects of which become small as $m_X^2$ increases.

The lightest of the exotic scalars is not stable because of the presence of the $Z_2$-breaking $M_2 X^3$ term in the scalar potential; this term will induce decays to pairs of SM gauge or Higgs bosons through a loop of the exotic scalars.

\subsection{GGM6}
\label{sec:ggm6}

The scalar potential for the GGM6 model can be written as
\begin{eqnarray}
	V(\Phi, X) &=& \frac{m^2_\Phi}{2} {\rm Tr} ( \Phi^\dagger \Phi )
	+ \frac{m^2_X}{2} {\rm Tr}( X^\dagger X)
	+ \lambda_1 \left[ {\rm Tr}( \Phi^\dagger \Phi) \right]^2 \nonumber  \\
	&& + \lambda_2 \left[ {\rm Tr}( X^\dagger X) \right]^2
	+ \lambda_3 {\rm Tr}( X^\dagger X X^\dagger X)
	+ \lambda_4 {\rm Tr} \left[ X^\dagger \widehat{T}^{1,i}_{5/2} X (\widehat{T}^{1,j}_{5/2})^\dagger \right] {\rm Tr} \left[ X^\dagger (\widehat{T}^{1,i}_{5/2})^\dagger X \widehat{T}^{1,j}_{5/2} \right]  \nonumber \\
	&& +  \lambda_5 {\rm Tr} \left[ X^\dagger \widehat{T}^{2,i}_{5/2} X (\widehat{T}^{2,j}_{5/2})^\dagger \right] {\rm Tr} \left[ X^\dagger (\widehat{T}^{2,i}_{5/2})^\dagger X \widehat{T}^{2,j}_{5/2} \right]
	+  \lambda_6 {\rm Tr} \left[ X^\dagger \widehat{T}^{2,i}_{5/2} X \right] {\rm Tr} \left[ X^\dagger (\widehat{T}^{2,i}_{5/2})^\dagger X \right] \nonumber \\
	&& +  \lambda_7 {\rm Tr} \left[ X^\dagger  X (\widehat{T}^{2,j}_{5/2})^\dagger \right] {\rm Tr}\left[ X^\dagger X  \widehat{T}^{2,j}_{5/2} \right] \nonumber\\
	&& + \lambda_8 {\rm Tr}( \Phi^\dagger \Phi){\rm Tr}( X^\dagger X) + \lambda_9 {\rm Tr}( \Phi^\dagger T_{1/2}^a \Phi T_{1/2}^b){\rm Tr}( X^\dagger T_{5/2}^a X T_{5/2}^b) \nonumber \\
	&& +  \lambda_{10} {\rm Tr}( X^\dagger \widehat{T}^{2,i}_{5/2} X  (\widehat{T}^{2,j}_{5/2})^\dagger) {\rm Tr}( \Phi^\dagger (\widehat{T}^{2,i}_{5/2,1/2})^\dagger X  \widehat{T}^{2,j}_{5/2,1/2}).
	\label{eq:GGM6pot}
\end{eqnarray}
The first line of this expression contains the two mass-squared terms and the doublet quartic coupling, just as in the GM model.  The next three lines contain the six linearly independent terms involving four powers of the $X_4$ field (in the GM model there are only two such terms).  The fifth line contains the two $\Phi^2 X^2$ terms; there are always only two ways to construct such terms, since the two doublets can be combined with total isospin zero or one.
The last line contains the term of the form $\Phi X^3$ that breaks the would-be $Z_2$ symmetry under which $X_6 \to -X_6$.  Again we have used the spherical tensors $\widehat T$ defined in Appendix~\ref{app:spherical} to write the potential in a compact form.

Minimizing the potential while assuming that custodial SU(2) is not spontaneously broken gives the two constraint equations,
\begin{eqnarray}
	0 = \frac{\partial V}{\partial v_{\phi}} &=&
		m_{\Phi}^2 v_{\phi} + 4 \lambda_1 v_{\phi}^3
		+ \frac{3}{4} \left[ 16 \lambda_8 + 35 \lambda_9 \right] v_{\phi} v_6^2
		+ 140 \sqrt{2} \lambda_{10} v_6^3, \nonumber \\
	0 = \frac{\partial V}{\partial v_6} &=&
		v_6 \left\{ 6 m_X^2 + \left[ 144 \lambda_2 + 24 \lambda_3 + 3675 \lambda_4
			+ 62720 \lambda_5 \right] v_6^2
		+ \frac{3}{4} \left[ 16 \lambda_8 + 35 \lambda_9 \right] v_{\phi}^2 +420 \sqrt{2} \lambda_{10} v_{\phi} v_6 \right\}.
	\label{eq:GGM6mincond1}
\end{eqnarray}
Again there are two phases: $v_6 = 0$ and $v_6 \neq 0$.  We first consider the case $v_6 \neq 0$; we will discuss the case $v_6 = 0$ below.

When $v_6$ is nonzero we can solve for $m_{\Phi}^2$ and $m_X^2$,
\begin{eqnarray}
	m^2_{\Phi} &=& -4 \lambda_1 v_{\phi}^2
	- \frac{3}{4} \left[ 16 \lambda_8 + 35 \lambda_9 \right] v_6^2
	- 140 \sqrt{2} \lambda_{10} \frac{v_6^3}{v_{\phi}}, \nonumber \\
	m^2_X &=& -\left[ 24 \lambda_2 + 4 \lambda_3 + \frac{1225}{2} \lambda_4 + \frac{31360}{3} \lambda_5 \right] v_6^2
	- \frac{1}{8} \left[ 16 \lambda_8 + 35 \lambda_9 \right] v_{\phi}^2
	- 70 \sqrt{2} \lambda_{10} v_{\phi} v_6.
	\label{eq:GGM6mincond}
\end{eqnarray}
The vevs are further constrained by the $W$ mass to obey
\begin{equation}
	v_{\phi}^2 + 70 v_6^2 = v^2,
	\label{eq:GGM6vevs}
\end{equation}
where $v^2$ is the SM Higgs vev given in Eq.~(\ref{eq:vevrelation}).

Using Eqs.~(\ref{eq:GGM6mincond}) to eliminate $m_{\Phi}^2$ and $m_X^2$, the masses of the physical states in the custodial elevenplet, nineplet, sevenplet, fiveplet, and triplet are given by
\begin{eqnarray}
	m_{11}^2 &=& -210 \left[5 \lambda_4 + 32 \lambda_5 \right] v_6^2
		- \frac{15}{2} \lambda_9 v_{\phi}^2
		- \frac{160 \sqrt{2}}{3} \lambda_{10} v_{\phi} v_6, \nonumber \\
	m_9^2 &=& 4 \left[ 2 \lambda_3 - 175 \lambda_4 - 3200 \lambda_5 \right] v_6^2
		- 5 \lambda_9 v_{\phi}^2
		- \frac{220 \sqrt{2}}{3} \lambda_{10} v_{\phi} v_6, \nonumber \\
	m_7^2 &=& -60 \left[ 7 \lambda_4 + 160 \lambda_5 \right] v_6^2
		- 3 \lambda_9 v_{\phi}^2
		- \frac{376 \sqrt{2}}{3} \lambda_{10} v_{\phi} v_6, \nonumber \\
	m_5^2 &=& \left[ 8 \lambda_3 + 238 \lambda_4 - \frac{21824}{3} \lambda_5
			+ 448 \lambda_6 + 448 \lambda_7 \right] v_6^2
		- \frac{3}{2} \lambda_9 v_{\phi}^2
		- 92 \sqrt{2} \lambda_{10} v_{\phi} v_6, \nonumber \\
	m_3^2 &=& -(v_{\phi}^2 + 70 v_6^2)
		\left[ \frac{1}{2} \lambda_9 + 2 \sqrt{2} \lambda_{10} \frac{v_6}{v_{\phi}} \right].
\end{eqnarray}
The elements of the custodial-singlet mass-squared matrix in the $(\phi^{0,r}, H_1^{\prime 0})$ basis are given by,
\begin{eqnarray}
	\mathcal{M}_{11}^2 &=& 8 \lambda_1 v_{\phi}^2
		- 140 \sqrt{2} \lambda_{10} \frac{v_6^3}{v_{\phi}}, \nonumber \\
	\mathcal{M}_{12}^2 &=& \frac{\sqrt{3}}{2 \sqrt{2}} \left[ 16 \lambda_8 + 35 \lambda_9 \right] v_{\phi} v_6 + 140 \sqrt{3} \lambda_{10} v_6^2, \nonumber \\
	\mathcal{M}_{22}^2 &=& 2 \left[ 24 \lambda_2 + 4 \lambda_3 + \frac{1225}{2} \lambda_4 + \frac{31360}{3} \lambda_5 \right] v_6^2
		+ 70 \sqrt{2} \lambda_{10} v_{\phi} v_6.
\end{eqnarray}
The mass eigenstates and mixing angle are defined as in Eq.~(\ref{eq:hmass}).
The compositions of the physical states are given explicitly in Appendix~\ref{app:states6}.

In the $v_6 \neq 0$ phase, the GGM6 model does \emph{not} possess a decoupling limit.  The easiest way to see this is to note that most of the masses of the exotic states have the form $\sum \lambda v^2$, which is bounded from above by the perturbativity of the quartic couplings and the $W$ mass constraint.  In this phase we again observe the strange feature that as $v_6 \to 0$, the mass of the lighter custodial-singlet scalar also goes to zero.  A lower bound on $v_6$ can thus be obtained from searches for a light custodial-singlet scalar.  This will again set a lower bound on the deviations of the couplings of the 125~GeV Higgs from their SM values.  This phase of the model could thus in principle be entirely ruled out by a combination of precision measurements of the 125~GeV Higgs boson couplings and searches for a light custodial-singlet scalar.

In the other phase $v_6 = 0$, the minimization condition $\partial V/\partial v_6 = 0$ holds automatically and $m_X^2$ cannot be eliminated from the potential.  In this case there is no mixing between the isospin doublet and the exotic scalars, and the additional scalars in $X_6$ decouple when $m_X^2 \gg v^2$.  In this phase the model is not interesting from the perspective of the LHC ``flat direction'' because $\kappa_V^h = \kappa_f^h = 1$; we include it here only for completeness.  The masses are now given by [we set $v_{\phi} = v$ as required by Eq.~(\ref{eq:GGM6vevs})]
\begin{eqnarray}
	m_{11}^2 &=& m_X^2 + 2 \lambda_8 v^2 - \frac{25}{8} \lambda_9 v^2, \nonumber \\
	m_9^2 &=& m_X^2 + 2 \lambda_8 v^2 - \frac{5}{8} \lambda_9 v^2, \nonumber \\
	m_7^2 &=& m_X^2 + 2 \lambda_8 v^2 + \frac{11}{8} \lambda_9 v^2, \nonumber \\
	m_5^2 &=& m_X^2 + 2 \lambda_8 v^2 + \frac{23}{8} \lambda_9 v^2, \nonumber \\
	m_3^2 &=& m_X^2 + 2 \lambda_8 v^2 + \frac{31}{8} \lambda_9 v^2, \nonumber \\
	m_H^2 &=& m_X^2 + 2 \lambda_8 v^2 + \frac{35}{8} \lambda_9 v^2, \nonumber \\
	m_h^2 &=& 8 \lambda_1 v^2,
\end{eqnarray}
where $h = \phi^{0,r}$ and $H = H_1^{\prime 0}$.  Note that the ordering of the masses of the exotic scalars is monotonic in their custodial SU(2) quantum number.  The lightest of the exotic scalars is not stable because of the presence of the $Z_2$-breaking $\lambda_{10} \Phi X^3$ term in the scalar potential; this term will induce decays to pairs of SM gauge or Higgs bosons through a loop of the exotic scalars.

\section{Conclusions}
\label{sec:conclusions}

In this paper we studied the generalizations of the GM model to higher isospin representations. We found that perturbative unitarity constraints restricted our considerations to just three generalized models. For each model we laid out the most general SU(2)$_L \times$SU(2)$_R$-invariant scalar potential and wrote down the masses and compositions of the scalars in the custodial-eigenstate basis.  This lays the groundwork for a comprehensive study of the theoretical constraints on the allowed parameter space of each model from perturbative unitarity and vacuum stability considerations, which is beyond the scope of the present paper.

We also surveyed the broad features of the phenomenology of each of the models by adapting existing analyses in the literature.  We showed how constraints on the GM model from $b \to s\gamma$ and the like-sign $WWjj$ cross section can be applied to the generalized GM models, and illustrated the resulting constraints on the maximum enhancement of the $hVV$ coupling.  We also obtained new constraints on the GM model and its generalizations at low custodial fiveplet masses from pair production of the custodial fiveplet states at the LHC and from a decay-mode-independent search for $Z H_5^0$ production at LEP.  At high custodial-fiveplet masses we obtained an additional new constraint from perturbative unitarity of the finite piece of the $VV \to VV$ scattering amplitudes, which limits the contribution of the larger multiplets to electroweak symmetry breaking when the custodial fiveplet is heavy.

The GM model and its three generalizations studied here, together with the septet model~\cite{Hisano:2013sn,Kanemura:2013mc}, comprise the \emph{complete set} of minimal weakly-coupled SM Higgs-sector extensions that preserve $\rho \equiv M_W^2/M_Z^2 \cos^2\theta_W = 1$ at tree level in a motivated way---i.e., due to custodial symmetry in the scalar potential or to a feature of the isospin and hypercharge quantum numbers of the new scalars.  They therefore provide us with a concrete framework in which to study scenarios in which the 125~GeV Higgs boson production rates in all channels are enhanced at the LHC, which can mask the presence of new, unobserved Higgs decay modes.   For example, the relationship between the $H_5 VV$ couplings and the maximum allowed enhancement of $\kappa_V^h$ given by the sum rule in Eq.~(\ref{eq:g52}) can be exploited to incorporate direct-search limits on $H_5$ production into the coupling fits for the 125~GeV Higgs boson.  These constraints, together with perturbative unitarity considerations, provide \emph{absolute} upper bounds on the 125~GeV Higgs boson's coupling to $WW$ and $ZZ$ based on the assumption that custodial SU(2) symmetry is preserved in the scalar sector.  Specifically, we find $\kappa_V^h < 2.36$, which is saturated at $m_5 = 97$~GeV.  This value is theoretically accessible only in the GGM5 and GGM6 models.

There are several clear directions in which our analysis can be extended:
\begin{itemize}
\item[$(i)$] The theoretical constraints on the generalized GM models will be very important in constraining the allowed enhancement of $\kappa_V^h$, especially when the additional states are heavy.  This has already been shown to be the case in the GM model~\cite{Hartling:2014zca,Hartling:2014aga}.  These constraints comprise perturbative unitarity of $2 \to 2$ scattering amplitudes involving the scalar quartic couplings, bounded-from-belowness of the scalar potential, and stability of the desired electroweak symmetry breaking and custodial symmetry preserving vacuum against decays into other vacua.

\item[$(ii)$] The parameter space of the generalized GM models will be further constrained by the oblique parameter $S$~\cite{Peskin:1991sw}.  This constraint has previously been studied in the GM model in Refs.~\cite{Chiang:2013rua,Kanemura:2013mc,Englert:2013zpa,Hartling:2014aga}.

\item[$(iii)$] Direct searches for the additional scalars at the LHC may put stringent constraints on the models containing larger isospin representations, due to the large multiplicity of states and their large weak charges.  Scalars in the custodial sevenplet and larger custodial multiplets will be pair-produced at the LHC through $s$-channel exchange of $W$ and $Z$ bosons and photons, and will then decay through $W$ or $Z$ emission to states in smaller custodial multiplets which can decay directly to pairs of SM particles.  Similar processes have been considered in the septet model and were found to be quite constraining even with the present LHC data~\cite{Alvarado:2014jva}.

\item[$(iv)$] Experimental searches for a very light custodial-singlet scalar in the GGM5 and GGM6 models could provide very interesting constraints on the parameter space of the $v_{5,6} \neq 0$ phases of these theories.  Because the second custodial-singlet scalar becomes very light at small values of $v_{5,6}$, these constraints would be complementary to the constraints obtained from searches for the custodial fiveplet states, which exclude large values of $v_{5,6}$.

\item[$(v)$] More than one bitriplet, biquartet, bipentet, and/or bisextet could be combined in a nonminimal custodial symmetry preserving Higgs sector extension.  This does not lead to a larger enhancement of the $hVV$ coupling compared to the generalized GM model containing only the largest of these representations because the maximum enhancement is fixed by the isospin of the largest additional representations as in Eqs.~(\ref{eq:kappaV}) and~(\ref{eq:A}).  However, adding additional SU(2)$_L \times$SU(2)$_R$ representations may allow direct-search constraints on the model parameter space to be loosened compared to the minimal generalized GM models considered here.  This is possible because such nonminimal extensions contain more than one custodial fiveplet, so that the $H_5$ coupling that appears in the sum rule in Eq.~(\ref{eq:g52}) can be shared among multiple states.  On the other hand, the proliferation of scalars in such extensions will likely make them even more vulnerable to exclusion by direct LHC searches.

\end{itemize}

Finally, we emphasize that many of the experimental constraints discussed here rely on the presence of custodial SU(2) symmetry in the scalar sector.  This assumption does not hold in the septet model.  A dedicated analysis based on the coupling relationships in that model is warranted.

\begin{acknowledgments}
We thank P.~Kalyniak for enlightening discussions about group theory, and C.-W.~Chiang, S.~Kanemura, and K.~Yagyu for providing the data points from Fig. 1 (left) of Ref.~\cite{Chiang:2014bia} that we used to make Fig.~\ref{fig:wwjj}.
H.E.L.\ also thanks J.~Gunion and H.~Haber for thought-provoking conversations and K.~Hartling and K.~Kunal for collaboration on the Georgi-Machacek model.
This work was supported by the Natural Sciences and Engineering Research Council of Canada.
\end{acknowledgments}

\appendix

\section{Generators of SU(2) in various representations}
\label{app:generators}

We list here for convenience the generators of SU(2) in various representations that are used to construct the Lagrangian terms for the generalized Georgi-Machacek models.

\subsection{Generators in the Cartesian basis}

To avoid confusion among the many indices, in the Cartesian basis we denote the gauge index $a$ by $(x, y, z)$ instead of the more common $(1,2,3)$.
In the $T = 1$ representation the generators are,
\begin{equation}
T^x_1 =
\left(
\begin{array}{ccc}
 0 & \frac{1}{\sqrt{2}} & 0 \\
 \frac{1}{\sqrt{2}} & 0 & \frac{1}{\sqrt{2}} \\
 0 & \frac{1}{\sqrt{2}} & 0 \\
\end{array}
\right), \qquad
T^y_1 =
\left(
\begin{array}{ccc}
 0 & -\frac{i}{\sqrt{2}} & 0 \\
 \frac{i}{\sqrt{2}} & 0 & -\frac{i}{\sqrt{2}} \\
 0 & \frac{i}{\sqrt{2}} & 0 \\
\end{array}
\right), \qquad
T^z_1 =
\left(
\begin{array}{ccc}
 1 & 0 & 0 \\
 0 & 0 & 0 \\
 0 & 0 & -1 \\
\end{array}
\right),
\end{equation}
and the combinations $T^{\pm} \equiv T^x \pm i T^y$ are,
\begin{equation}
T^+_1 =
\left(
\begin{array}{ccc}
 0 & \sqrt{2} & 0 \\
 0 & 0 & \sqrt{2} \\
 0 & 0 & 0 \\
\end{array}
\right)
, \qquad
T^-_1 =
\left(
\begin{array}{ccc}
 0 & 0 & 0 \\
 \sqrt{2} & 0 & 0 \\
 0 & \sqrt{2} & 0 \\
\end{array}
\right).
\end{equation}

In the $T = 3/2$ representation the generators are,
\begin{equation}
T^x_\frac{3}{2} =
\left(
\begin{array}{cccc}
 0 & \frac{\sqrt{3}}{2} & 0 & 0 \\
 \frac{\sqrt{3}}{2} & 0 & 1 & 0 \\
 0 & 1 & 0 & \frac{\sqrt{3}}{2} \\
 0 & 0 & \frac{\sqrt{3}}{2} & 0 \\
\end{array}
\right), \qquad
T^y_\frac{3}{2} =
\left(
\begin{array}{cccc}
 0 & -\frac{i \sqrt{3}}{2} & 0 & 0 \\
 \frac{i \sqrt{3}}{2} & 0 & -i & 0 \\
 0 & i & 0 & -\frac{i \sqrt{3}}{2} \\
 0 & 0 & \frac{i \sqrt{3}}{2} & 0 \\
\end{array}
\right), \qquad
T^z_\frac{3}{2} =
\left(
\begin{array}{cccc}
 \frac{3}{2} & 0 & 0 & 0 \\
 0 & \frac{1}{2} & 0 & 0 \\
 0 & 0 & -\frac{1}{2} & 0 \\
 0 & 0 & 0 & -\frac{3}{2} \\
\end{array}
\right),
\end{equation}
\begin{equation}
T^+_\frac{3}{2} =
\left(
\begin{array}{cccc}
 0 & \sqrt{3} & 0 & 0 \\
 0 & 0 & 2 & 0 \\
 0 & 0 & 0 & \sqrt{3} \\
 0 & 0 & 0 & 0 \\
\end{array}
\right)
, \qquad
T^-_\frac{3}{2} =
\left(
\begin{array}{cccc}
 0 & 0 & 0 & 0 \\
 \sqrt{3} & 0 & 0 & 0 \\
 0 & 2 & 0 & 0 \\
 0 & 0 & \sqrt{3} & 0 \\
\end{array}
\right).
\end{equation}

In the $T = 2$ representation the generators are,
\begin{equation}
T^x_2 =
\left(
\begin{array}{ccccc}
 0 & 1 & 0 & 0 & 0 \\
 1 & 0 & \sqrt{\frac{3}{2}} & 0 & 0 \\
 0 & \sqrt{\frac{3}{2}} & 0 & \sqrt{\frac{3}{2}} & 0 \\
 0 & 0 & \sqrt{\frac{3}{2}} & 0 & 1 \\
 0 & 0 & 0 & 1 & 0 \\
\end{array}
\right), \qquad
T^y_2 =
\left(
\begin{array}{ccccc}
 0 & -i & 0 & 0 & 0 \\
 i & 0 & -i \sqrt{\frac{3}{2}} & 0 & 0 \\
 0 & i \sqrt{\frac{3}{2}} & 0 & -i \sqrt{\frac{3}{2}} & 0 \\
 0 & 0 & i \sqrt{\frac{3}{2}} & 0 & -i \\
 0 & 0 & 0 & i & 0 \\
\end{array}
\right), \qquad
T^z_2 =
\left(
\begin{array}{ccccc}
 2 & 0 & 0 & 0 & 0 \\
 0 & 1 & 0 & 0 & 0 \\
 0 & 0 & 0 & 0 & 0 \\
 0 & 0 & 0 & -1 & 0 \\
 0 & 0 & 0 & 0 & -2 \\
\end{array}
\right),
\end{equation}
\begin{equation}
T^+_2 =
\left(
\begin{array}{ccccc}
 0 & 2 & 0 & 0 & 0 \\
 0 & 0 & \sqrt{6} & 0 & 0 \\
 0 & 0 & 0 & \sqrt{6} & 0 \\
 0 & 0 & 0 & 0 & 2 \\
 0 & 0 & 0 & 0 & 0 \\
\end{array}
\right)
, \qquad
T^-_2 =
\left(
\begin{array}{ccccc}
 0 & 0 & 0 & 0 & 0 \\
 2 & 0 & 0 & 0 & 0 \\
 0 & \sqrt{6} & 0 & 0 & 0 \\
 0 & 0 & \sqrt{6} & 0 & 0 \\
 0 & 0 & 0 & 2 & 0 \\
\end{array}
\right).
\end{equation}

Finally, in the $T = 5/2$ representation the generators are,
\begin{eqnarray}
T^x_\frac{5}{2} &=&
\left(
\begin{array}{cccccc}
 0 & \frac{\sqrt{5}}{2} & 0 & 0 & 0 & 0 \\
 \frac{\sqrt{5}}{2} & 0 & \sqrt{2} & 0 & 0 & 0 \\
 0 & \sqrt{2} & 0 & \frac{3}{2} & 0 & 0 \\
 0 & 0 & \frac{3}{2} & 0 & \sqrt{2} & 0 \\
 0 & 0 & 0 & \sqrt{2} & 0 & \frac{\sqrt{5}}{2} \\
 0 & 0 & 0 & 0 & \frac{\sqrt{5}}{2} & 0 \\
\end{array}
\right), \qquad
T^y_\frac{5}{2} =
\left(
\begin{array}{cccccc}
 0 & -\frac{i \sqrt{5}}{2} & 0 & 0 & 0 & 0 \\
 \frac{i \sqrt{5}}{2} & 0 & -i \sqrt{2} & 0 & 0 & 0 \\
 0 & i \sqrt{2} & 0 & -\frac{3 i}{2} & 0 & 0 \\
 0 & 0 & \frac{3 i}{2} & 0 & -i \sqrt{2} & 0 \\
 0 & 0 & 0 & i \sqrt{2} & 0 & -\frac{i \sqrt{5}}{2} \\
 0 & 0 & 0 & 0 & \frac{i \sqrt{5}}{2} & 0 \\
\end{array}
\right), \nonumber \\
&& T^z_\frac{5}{2} =
\left(
\begin{array}{cccccc}
 \frac{5}{2} & 0 & 0 & 0 & 0 & 0 \\
 0 & \frac{3}{2} & 0 & 0 & 0 & 0 \\
 0 & 0 & \frac{1}{2} & 0 & 0 & 0 \\
 0 & 0 & 0 & -\frac{1}{2} & 0 & 0 \\
 0 & 0 & 0 & 0 & -\frac{3}{2} & 0 \\
 0 & 0 & 0 & 0 & 0 & -\frac{5}{2} \\
\end{array}
\right),
\end{eqnarray}
\begin{equation}
T^+_\frac{5}{2} =
\left(
\begin{array}{cccccc}
 0 & \sqrt{5} & 0 & 0 & 0 & 0 \\
 0 & 0 & 2 \sqrt{2} & 0 & 0 & 0 \\
 0 & 0 & 0 & 3 & 0 & 0 \\
 0 & 0 & 0 & 0 & 2 \sqrt{2} & 0 \\
 0 & 0 & 0 & 0 & 0 & \sqrt{5} \\
 0 & 0 & 0 & 0 & 0 & 0 \\
\end{array}
\right)
, \qquad
T^-_\frac{5}{2} =
\left(
\begin{array}{cccccc}
 0 & 0 & 0 & 0 & 0 & 0 \\
 \sqrt{5} & 0 & 0 & 0 & 0 & 0 \\
 0 & 2 \sqrt{2} & 0 & 0 & 0 & 0 \\
 0 & 0 & 3 & 0 & 0 & 0 \\
 0 & 0 & 0 & 2 \sqrt{2} & 0 & 0 \\
 0 & 0 & 0 & 0 & \sqrt{5} & 0 \\
\end{array}
\right).
\end{equation}

\subsection{Generators in the spherical basis}
\label{app:spherical}

The spherical tensors and mixed spherical tensors are useful when combining pairs of scalar fields into particular representations of SU(2)$_L \times$SU(2)$_R$.
We use the notation $\widehat{T}^{j,i}_r$ to denote the $i$-th spherical tensor of rank $j$ constructed from the SU(2) generators in the spherical basis $\left( -\frac{1}{\sqrt{2}} T^+_r , T^z_r, \frac{1}{\sqrt{2}} T^-_r \right)$, where $r$ denotes the representation and $T^z \equiv T^3$.
Thus, the rank-1 spherical tensors in representation $r$ are just,
\begin{eqnarray}
\widehat{T}^{1,1}_r &=&  -\frac{1}{\sqrt{2}} T^+_r, \qquad
\widehat{T}^{1,0}_r =  T^z_r, \qquad
\widehat{T}^{1,-1}_r = \frac{1}{\sqrt{2}} T^-_r.
\end{eqnarray}

The rank-2 spherical tensors in representation $r$ are given by,
\begin{eqnarray}
	\widehat{T}^{2,2}_r &=& \widehat{T}^{1,1}_r \widehat{T}^{1,1}_r , \nonumber \\
	\widehat{T}^{2,1}_r &=&  \frac{1}{\sqrt{2}} \left( \widehat{T}^{1,1}_r  \widehat{T}^{1,0}_r  + \widehat{T}^{1,0}_r  \widehat{T}^{1,1}_r \right), \nonumber \\
	\widehat{T}^{2,0}_r &=& \frac{1}{\sqrt{6}} \left( \widehat{T}^{1,1}_r \widehat{T}^{1,-1}_r + \widehat{T}^{1,-1}_r \widehat{T}^{1,1}_r + 2 \widehat{T}^{1,0}_r \widehat{T}^{1,0}_r \right), \nonumber \\
	\widehat{T}^{2,-1}_r &=& \frac{1}{\sqrt{2}} \left( \widehat{T}^{1,-1}_r  \widehat{T}^{1,0}_r  + \widehat{T}^{1,0}_r  \widehat{T}^{1,-1}_r \right), \nonumber \\
	\widehat{T}^{2,-2}_r &=& \widehat{T}^{1,-1}_r \widehat{T}^{1,-1}_r.
\end{eqnarray}

For a representation $r = j_1$, each of the spherical tensors is a $(2j_1+1) \times (2j_1+1)$ matrix, whose indices we can denote as $m_1$ and $m_2$. Then, the spherical tensor of rank $j$ can be shown to be simply related to the Clebsch-Gordan coefficients $C^{j,m}_{j_1,m_1^\prime, j_1,m_2}$ as,
\begin{equation}
	\left(\widehat{T}^{j,m}_{j_1} \right)_{m_1,m_2}
	\propto \mathcal{C}^{j_1}_{m_1,m_1^\prime} C^{j,m}_{j_1,m_1^\prime, j_1,m_2},
\end{equation}
where $\mathcal{C}^{j_1}_{m_1,m_1^\prime}$ is the charge-conjugation operator defined as
\begin{equation}
	\mathcal{C}^{j_1}_{m_1,m_1^\prime} = (-1)^{m_1-j_1} \delta_{m_1,-m_1^{\prime}}.
\end{equation}
This charge-conjugation operator is a $(2j_1+1) \times (2j_1+1)$ antidiagonal matrix with $+1$ in the upper right-hand corner and alternating signs $(+1, -1, +1, \ldots)$ down the antidiagonal.

We can easily generalize this to produce ``mixed'' spherical tensors, which are used in the scalar potentials for the GGM4 and GGM6 models in Secs.~\ref{sec:ggm4} and \ref{sec:ggm6}, respectively.
The $m$-th mixed spherical tensors of rank $j$ constructed from representations $j_1$ and $j_2$ are given by,
\begin{equation}
	\left(\widehat{T}^{j,m}_{j_1,j_2}  \right)_{m_1,m_2}  = \mathcal{C}^{j_1}_{m_1,m_1^\prime} C^{j,m}_{j_1,m_1^\prime, j_2,m_2}.
\end{equation}

In the GGM4 model we use
\begin{eqnarray}
	(\widehat{T}^{1,1}_{3/2,1/2})^\dagger &=& \left(\begin{array}{cccc}
		 -\frac{\sqrt{3}}{2} & 0 & 0 & 0 \\
		 0 & -\frac{1}{2} & 0 & 0 \\
		\end{array} \right), \nonumber \\
	(\widehat{T}^{1,0}_{3/2,1/2})^\dagger &=& \left(\begin{array}{cccc}
		 0 & -\frac{1}{\sqrt{2}} & 0 & 0 \\
 		0 & 0 & -\frac{1}{\sqrt{2}} & 0 \\
		\end{array} \right), \nonumber \\
	(\widehat{T}^{1,-1}_{3/2,1/2})^\dagger &=& \left(\begin{array}{cccc}
		 0 & 0 & -\frac{1}{2} & 0 \\
		 0 & 0 & 0 & -\frac{\sqrt{3}}{2} \\
		\end{array} \right).
\end{eqnarray}

In the GGM6 model we use
\begin{eqnarray}
	(\widehat{T}^{2,2}_{5/2,1/2})^\dagger &=& \left( \begin{array}{cccccc}
		 -\sqrt{\frac{5}{6}} & 0 & 0 & 0 & 0 & 0 \\
		 0 & -\frac{1}{\sqrt{6}} & 0 & 0 & 0 & 0 \\
		\end{array} \right), \nonumber \\
	(\widehat{T}^{2,1}_{5/2,1/2})^\dagger &=& \left( \begin{array}{cccccc}
		 0 & -\sqrt{\frac{2}{3}} & 0 & 0 & 0 & 0 \\
		 0 & 0 & -\frac{1}{\sqrt{3}} & 0 & 0 & 0 \\
		\end{array} \right), \nonumber \\
	(\widehat{T}^{2,0}_{5/2,1/2})^\dagger &=& \left( \begin{array}{cccccc}
		 0 & 0 & -\frac{1}{\sqrt{2}} & 0 & 0 & 0 \\
		 0 & 0 & 0 & -\frac{1}{\sqrt{2}} & 0 & 0 \\
		\end{array} \right), \nonumber \\
	(\widehat{T}^{2,-1}_{5/2,1/2})^\dagger &=& \left( \begin{array}{cccccc}
		 0 & 0 & 0 & -\frac{1}{\sqrt{3}} & 0 & 0 \\
		 0 & 0 & 0 & 0 & -\sqrt{\frac{2}{3}} & 0 \\
		\end{array} \right), \nonumber \\
	(\widehat{T}^{2,-2}_{5/2,1/2})^\dagger &=& \left( \begin{array}{cccccc}
		 0 & 0 & 0 & 0 & -\frac{1}{\sqrt{6}} & 0 \\
		 0 & 0 & 0 & 0 & 0 & -\sqrt{\frac{5}{6}} \\
		\end{array} \right).
\end{eqnarray}

\section{Explicit notation for the scalars}
\label{app:states}

\subsection{GGM4}
\label{app:states4}

The biquartet can be written as
\begin{equation}
	X_4 = \left( \begin{array}{cccc}
		\psi_3^{0*} & -\psi_1^{-*} & \psi_1^{++} & \psi_3^{+3} \\
		-\psi_3^{+*} & \psi_1^{0*} & \psi_1^+ & \psi_3^{++} \\
		\psi_3^{++*} & -\psi_1^{+*} & \psi_1^0 & \psi_3^+ \\
		-\psi_3^{+3*} & \psi_1^{++*} & \psi_1^- & \psi_3^0 \end{array} \right),
\end{equation}
where the subscripts denote the hypercharge of the two SU(2)$_L$ quartets.  After electroweak symmetry breaking the neutral states decompose according to
\begin{equation}
	\psi_j^0 \to v_4 + (\psi_j^{0,r} + i \psi_j^{0,i})/\sqrt{2}, \qquad j = 1,3.
\end{equation}

The biquartet decomposes into a singlet $H_1^{\prime}$, triplet $H_3^{\prime}$, fiveplet $H_5$, and sevenplet $H_7$ under custodial SU(2).  (The custodial singlet and triplet subsequently mix with the corresponding states from the doublet to form mass eigenstates.)  The custodial singlet and triplet can be obtained from general expressions given in Ref.~\cite{Logan:1999if}:
\begin{eqnarray}
	H_1^{\prime 0} &=& (\psi_1^{0,r} + \psi_3^{0,r})/\sqrt{2}, \nonumber \\
	H_3^{\prime 0} &=& (\psi_1^{0,i} + 3 \psi_3^{0,i})/\sqrt{10}, \nonumber \\
	H_3^{\prime +} &=& (-\sqrt{3} \psi_1^{-*} + 2 \psi_1^+ + \sqrt{3} \psi_3^+)/\sqrt{10}.
\end{eqnarray}
The custodial fiveplet and sevenplet are given by:
\begin{eqnarray}
    H_5^{++} &=& (\psi_1^{++} + \psi_3^{++})/\sqrt{2}, \nonumber \\
	H_5^{+} &=& (\psi_1^{-*} +  \psi_3^{+})/\sqrt{2}, \nonumber \\
	H_5^{0} &=& (\psi_3^{0,r}  - \psi_1^{0,r}  )/\sqrt{2}, \nonumber \\
    H_7^{+3} &=& \psi_3^{+3},  \nonumber \\
    H_7^{++} &=& (\psi_3^{++} - \psi_1^{++})/\sqrt{2}, \nonumber \\
    H_7^{+} &=& (\psi_3^{+}-\psi_1^{-*} - \sqrt{3} \psi_1^+ )/\sqrt{5}, \nonumber \\
	H_7^{0} &=& (\psi_3^{0,i} -3\psi_1^{0,i}  )/\sqrt{10}.
\end{eqnarray}

\subsection{GGM5}
\label{app:states5}

The bipentet can be written as
\begin{equation}
	X_5 = \left( \begin{array}{ccccc}
		\pi_4^{0*} & -\pi_2^{-*} & \pi_0^{++} & \pi_2^{+3} & \pi_4^{+4} \\
		-\pi_4^{+*} & \pi_2^{0*} & \pi_0^+ & \pi_2^{++} & \pi_4^{+3} \\
		\pi_4^{++*} & -\pi_2^{+*} & \pi_0^0 & \pi_2^+ & \pi_4^{++} \\
		-\pi_4^{+3*} & \pi_2^{++*} & -\pi_0^{+*} & \pi_2^0 & \pi_4^+ \\
		\pi_4^{+4*} & -\pi_2^{+3*} & \pi_0^{++*} & \pi_2^- & \pi_4^0
		\end{array} \right),
\end{equation}
where the subscripts denote the hypercharge of the SU(2)$_L$ pentets.  $\pi_2$ and $\pi_4$ are complex pentets while $\pi_0$ is a real pentet.  After electroweak symmetry breaking the neutral states decompose according to
\begin{eqnarray}
	\pi_j^0 &\to& v_5 + (\pi_j^{0,r} + i \pi_j^{0,i})/\sqrt{2}, \qquad j = 2,4, \nonumber \\
	\pi_0^0 &\to& v_5 + \pi_0^0,
\end{eqnarray}
where $\pi_0^0$ is already a real field.

The bipentet decomposes into a singlet $H_1^{\prime}$, triplet $H_3^{\prime}$, fiveplet $H_5$, sevenplet $H_7$, and nineplet $H_9$ under custodial SU(2).  (The custodial singlet and triplet subsequently mix with the corresponding states from the doublet to form mass eigenstates.)  The custodial singlet and triplet can be obtained from general expressions given in Ref.~\cite{Logan:1999if}:
\begin{eqnarray}
	H_1^{\prime 0} &=& (\pi_0^0 + \sqrt{2} \pi_2^{0,r} + \sqrt{2} \pi_4^{0,r})/\sqrt{5}, \nonumber \\
	H_3^{\prime 0} &=& (\pi_2^{0,i} + 2 \pi_4^{0,i})/\sqrt{5}, \nonumber \\
	H_3^{\prime +} &=& (-\sqrt{2} \pi_2^{-*} + \sqrt{3} \pi_0^+ + \sqrt{3} \pi_2^+ + \sqrt{2} \pi_4^+)/\sqrt{10}.
\end{eqnarray}
The custodial fiveplet, sevenplet and nineplet are given by:
\begin{eqnarray}
	H_5^{++} &=& (\sqrt{2} \pi_0^{++} + \sqrt{2} \pi_4^{++} +\sqrt{3} \pi_2^{++})/\sqrt{7}, \nonumber \\
	H_5^{+} &=& ( \sqrt{6} \pi_2^{-*} -  \pi_0^{+} +  \pi_2^{+} +\sqrt{6}\pi_4^{+} )/\sqrt{14}, \nonumber \\
	H_5^{0} &=& ( 2 \pi_4^{0,r}-\sqrt{2}\pi_0^{0} -  \pi_2^{0,r} )/\sqrt{7}, \nonumber \\
    H_7^{+3} &=& ( \pi_2^{+3} + \pi_4^{+3})/\sqrt{2}, \nonumber \\
	H_7^{++} &=& (\pi_4^{++}-\pi_0^{++}  )/\sqrt{2}, \nonumber \\
	H_7^{+} &=& ( \sqrt{3}\pi_4^{+}- \sqrt{3} \pi_2^{-*} - \sqrt{2}  \pi_0^{+} -  \sqrt{2} \pi_2^{+} )/\sqrt{10}, \nonumber \\
	H_7^{0} &=& (\pi_4^{0,i}- 2\pi_2^{0,i} )/\sqrt{5}, \nonumber \\
    H_9^{+4} &=& \pi_4^{+4}, \nonumber \\
    H_9^{+3} &=& ( \pi_4^{+3} -  \pi_2^{+3} )/\sqrt{2}, \nonumber \\
    H_9^{++} &=& (\sqrt{3}\pi_0^{++}+\sqrt{3} \pi_4^{++} - 2 \sqrt{2} \pi_2^{++} )/\sqrt{14}, \nonumber \\
	H_9^{+} &=& (  \pi_2^{-*} + \sqrt{6}  \pi_0^{+} - \sqrt{6} \pi_2^{+} + \pi_4^{+} )/\sqrt{14}, \nonumber \\
	H_9^{0} &=& (3\sqrt{2} \pi_0^0 - 4 \pi_2^{0,r} + \pi_4^{0,r})/\sqrt{35}.
\end{eqnarray}

\subsection{GGM6}
\label{app:states6}

The bisextet can be written as
\begin{equation}
	X_6 = \left( \begin{array}{ccc ccc}
	\zeta_5^{0*} & -\zeta_3^{-*} & \zeta_1^{--*} & \zeta_1^{+3} & \zeta_3^{+4} & \zeta_5^{+5} \\
	-\zeta_5^{+*} & \zeta_3^{0*} & -\zeta_1^{-*} & \zeta_1^{++} & \zeta_3^{+3} & \zeta_5^{+4} \\
	\zeta_5^{++*} & -\zeta_3^{+*} & \zeta_1^{0*} & \zeta_1^{+} & \zeta_3^{++} & \zeta_5^{+3} \\
	-\zeta_5^{+3*} & \zeta_3^{++*} & -\zeta_1^{+*} & \zeta_1^0 & \zeta_3^{+} & \zeta_5^{++} \\
	\zeta_5^{+4*} & -\zeta_3^{+3*} & \zeta_1^{++*} & \zeta_1^- & \zeta_3^0 & \zeta_5^+ \\
	-\zeta_5^{+5*} & \zeta_3^{+4*} & -\zeta_1^{+3*} & \zeta_1^{--} & \zeta_3^- &\zeta_5^0
	\end{array} \right),
\end{equation}
where the subscripts denote the hypercharge of the three SU(2)$_L$ sextets.  After electroweak symmetry breaking the neutral states decompose according to
\begin{equation}
	\zeta_j^0 \to v_6 + (\zeta_j^{0,r} + i \zeta_j^{0,i})/\sqrt{2}, \qquad j = 1,3,5.
\end{equation}

The bisextet decomposes into a singlet $H_1^{\prime}$, triplet $H_3^{\prime}$, fiveplet $H_5$, sevenplet $H_7$, nineplet $H_9$, and elevenplet $H_{11}$ under custodial SU(2).  (The custodial singlet and triplet subsequently mix with the corresponding states from the doublet to form mass eigenstates.)  The custodial singlet and triplet can be obtained from general expressions given in Ref.~\cite{Logan:1999if}:
\begin{eqnarray}
	H_1^{\prime 0} &=& (\zeta_1^{0,r} + \zeta_3^{0,r} + \zeta_5^{0,r})/\sqrt{3}, \nonumber \\
	H_3^{\prime 0} &=& (\zeta_1^{0,i} + 3 \zeta_3^{0,i} + 5 \zeta_5^{0,i})/\sqrt{35}, \nonumber \\
	H_3^{\prime +} &=& (-\sqrt{5} \zeta_3^{-*} - \sqrt{8} \zeta_1^{-*} + 3 \zeta_1^+ + \sqrt{8} \zeta_3^+ + \sqrt{5} \zeta_5^+)/\sqrt{35}.
\end{eqnarray}
The custodial fiveplet, sevenplet, nineplet and elevenplet are given by:
\begin{eqnarray}
	H_{5}^{++} &=& (\sqrt{5}\zeta_1^{--*} +3 \zeta_1^{++} +3 \zeta_3^{++} +\sqrt{5} \zeta_5^{++})/\sqrt{28}, \nonumber \\
	H_{5}^{+} &=& (2 \zeta_1^{-*} +  \sqrt{10}\zeta_3^{-*} +2 \zeta_3^{+} +\sqrt{10}\zeta_5^+ )/\sqrt{28}, \nonumber \\
	H_{5}^{0} &=& ( 5\zeta_5^{0,r}- 4\zeta_1^{0,r} - \zeta_3^{0,r} )/\sqrt{42}, \nonumber \\
    H_{7}^{+3} &=& (\sqrt{10} \zeta_1^{+3} +\sqrt{10}  \zeta_5^{+3} + 4 \zeta_3^{+3})/6, \nonumber \\
	H_{7}^{++} &=& (  \zeta_3^{++} +\sqrt{5} \zeta_5^{++}-\sqrt{5}\zeta_1^{--*} - \zeta_1^{++})/\sqrt{12}, \nonumber \\
	H_{7}^{+} &=& ( \zeta_1^{-*} -  \sqrt{10}\zeta_3^{-*}- \sqrt{8} \zeta_1^{+}  -\zeta_3^{+} +\sqrt{10}\zeta_5^+ )/\sqrt{30}, \nonumber \\
	H_{7}^{0} &=& (  5\zeta_5^{0,i}-4\zeta_1^{0,i} - 7 \zeta_3^{0,i})/\sqrt{90}, \nonumber \\
	H_{9}^{+4} &=& (\zeta_5^{+4} +  \zeta_3^{+4})/\sqrt{2}, \nonumber \\
	H_{9}^{+3} &=& (\zeta_5^{+3} -\zeta_1^{+3}   )/\sqrt{2}, \nonumber \\
	H_{9}^{++} &=& (3\zeta_1^{--*} - \sqrt{5}\zeta_1^{++} -\sqrt{5} \zeta_3^{++} +3 \zeta_5^{++})/\sqrt{28}, \nonumber \\
	H_{9}^{+} &=& (2\zeta_5^+ -\sqrt{10} \zeta_1^{-*} +  2\zeta_3^{-*}  -\sqrt{10}\zeta_3^{+}  )/\sqrt{28}, \nonumber \\
	H_{9}^{0} &=& (2\zeta_1^{0,r} - 3 \zeta_3^{0,r} + \zeta_5^{0,r})/\sqrt{14}, \nonumber \\
	H_{11}^{+5} &=& \zeta_5^{+5}, \nonumber \\
	H_{11}^{+4} &=& (\zeta_5^{+4} -  \zeta_3^{+4})/\sqrt{2}, \nonumber \\
	H_{11}^{+3} &=& (\sqrt{2} \zeta_1^{+3} - \sqrt{5} \zeta_3^{+3} + \sqrt{2} \zeta_5^{+3})/3 , \nonumber \\
	H_{11}^{++} &=& (-\zeta_1^{--*} + \sqrt{5}\zeta_1^{++} -\sqrt{5} \zeta_3^{++} + \zeta_5^{++})/\sqrt{12}, \nonumber \\
	H_{11}^{+} &=& (\sqrt{10} \zeta_1^{-*} -  \zeta_3^{-*} + 2\sqrt{5} \zeta_1^{+} -\sqrt{10}\zeta_3^{+} +\zeta_5^+ )/\sqrt{42}, \nonumber \\
	H_{11}^{0} &=& (10 \zeta_1^{0,i} - 5 \zeta_3^{0,i} + \zeta_5^{0,i})/\sqrt{126}.
\end{eqnarray}


\end{document}